\documentclass[11pt,a4paper]{article}
\pdfoutput=1
\usepackage{jheppub}
\usepackage{graphicx} 
\usepackage{epstopdf} 
\usepackage{dcolumn} 
\usepackage{subcaption}
\usepackage[export]{adjustbox}
\usepackage{xcolor} 
\usepackage{amsmath,amssymb} 
\usepackage{hyperref} 
\usepackage[utf8]{inputenc} 
\usepackage[english]{babel} 
\usepackage[mathlines]{lineno} 
\usepackage{blindtext} 
\usepackage{ifthen} 
\usepackage{orcidlink} 
\usepackage{physics}
\usepackage[normalem]{ulem}
\usepackage{multirow}
\usepackage{makecell}

\graphicspath{{figures/}} 

\newboolean{articletitles}
\setboolean{articletitles}{true} 

\input{belle2-symbols}

\begin{document}

\vspace*{-3\baselineskip}
\resizebox{!}{2cm}{\includegraphics{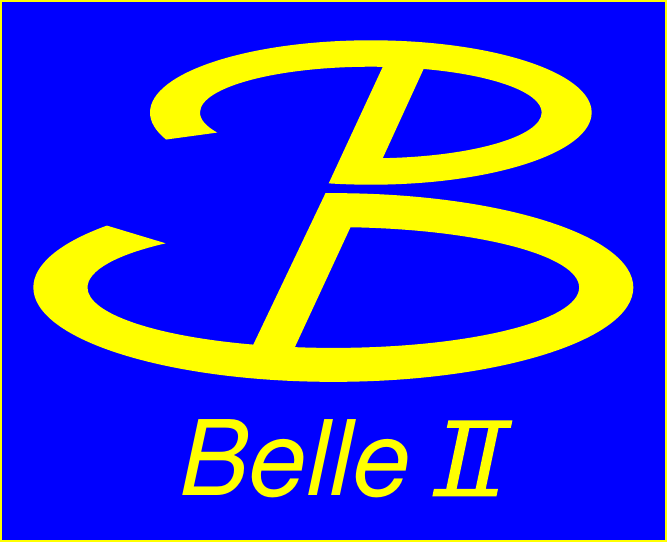}}
\begin{flushright}
Belle II preprint: 2026-008\\
KEK preprint: 2026-4
\end{flushright}

\title{
Search for the $\boldsymbol{B^0 \to \KS \tau^+\tau^-}$ decay
}
\collaboration{The Belle and Belle II Collaborations}
  \author{M.~Abumusabh\,\orcidlink{0009-0004-1031-5425},} 
  \author{I.~Adachi\,\orcidlink{0000-0003-2287-0173},} 
  \author{A.~Aggarwal\,\orcidlink{0000-0002-5623-3896},} 
  \author{Y.~Ahn\,\orcidlink{0000-0001-6820-0576},} 
  \author{H.~Aihara\,\orcidlink{0000-0002-1907-5964},} 
  \author{M.~Akdag\,\orcidlink{0009-0004-3728-1077},} 
  \author{N.~Akopov\,\orcidlink{0000-0002-4425-2096},} 
  \author{S.~Alghamdi\,\orcidlink{0000-0001-7609-112X},} 
  \author{M.~Alhakami\,\orcidlink{0000-0002-2234-8628},} 
  \author{A.~Aloisio\,\orcidlink{0000-0002-3883-6693},} 
% ...

\abstract{

We present the first search for $B^0 \to \KS \tau^+\tau^-$ decays. We look for signal decays in $B^0\bar B^0$ events produced in asymmetric-energy electron-positron collisions. This work uses samples from the Belle and Belle~II detectors, comprising 1.16 billion $\Upsilon(4S)$ events.
In $\Upsilon(4S)\to B^0\bar{B}^0$ decays, the non-signal $\bar{B}^0$ meson is fully reconstructed in a hadronic channel.
For the signal $B^0$ meson, $\tau$-lepton decays into final states with a single charged particle are selected. We observe no evidence for the signal and set an upper limit on the branching fraction $\mathcal{B}(B^0\to \KS \tau^+\tau^-) < 8.3 \times 10^{-4}$ at the 90\% confidence level. Combining this with the recent measurement of the isospin-partner decay $B^+\to K^+\tau^+\tau^-$, we determine an upper limit $\mathcal{B}(B\to K\tau^+\tau^-) < 5.4\times10^{-4}$ at the 90\% confidence level.



}

\maketitle
\flushbottom


\section{Introduction} \label{sec:intro}

The semileptonic $b\to s\tau^+\tau^-$ decays are flavour-changing neutral-current processes involving third-generation quarks and leptons. In the Standard Model (SM) they are predicted to be rare, with branching fractions of $\mathcal{O}(10^{-7})$~\cite{Parrott}. 
Their decay rates are sensitive to heavy degrees of freedom beyond the Standard Model (BSM) but
their observation is experimentally challenging due to small branching fractions and the presence of multiple neutrinos in the final state.
Measurements in the related $b \to s\nu\bar{\nu}$, $b\to s\mu^+\mu^-$ and $b\to c\tau^-\bar{\nu}_{\tau}$ transitions~\cite{btoknunu,Chen:2024jlj,f00hflav} show some disagreement with the SM predictions. These discrepancies can be accommodated in BSM scenarios with new heavy mediators, such as leptoquarks or $Z'$ bosons, that violate lepton-flavour universality~\cite{anom1,anom2}.  
The $b\to s\tau^+\tau^-$ decay modes can be enhanced by up to three orders of magnitude through couplings to such mediators~\cite{Capdevila,Allwicher,Bause}, making them a valuable probe of the interplay between the neutral- and charged-current sectors.

Recent searches have set upper limits on the branching fractions of $B^+\to K^+\tau^+\tau^-$ and $B^0\to K^{*0}\tau^+\tau^-$ decays\footnote{Charge-conjugate processes are implied throughout this paper.} at the level of a few times $10^{-4}$~\cite{butokplustautauB1B2,LHCb:2025lcw}, significantly improving upon earlier results~\cite{bstautau,butokplustautau,bdtokstartautau,bdtokstartautauB2}. However, the sensitivity to different classes of BSM mediators depends on the spin-parity of the hadronic final state~\cite{Bause}. Decays involving pseudoscalar mesons are less well explored than their vector meson counterparts and therefore offer a promising testing ground.

This work presents the first search for $B^0 \to \KS \tau^+\tau^-$ decays. We use electron-positron collision data at the $\Upsilon(4S)$ resonance from the Belle and Belle~II experiments with integrated luminosities of $711 \ \mathrm{fb}^{-1}$ and $365 \ \mathrm{fb}^{-1}$, respectively.
With this $\Upsilon(4S)\to B^0 \bar {B}^0$ dataset, the SM signal branching fraction $\mathcal{B}(B^0\to \KS\tau^+\tau^-) =0.5\times\mathcal{B}(B^0\to K^0\tau^+\tau^-)=0.5\times1.55\times10^{-7}$~\cite{Parrott} corresponds to about 100 signal events. For comparison, the backgrounds are $\mathcal{O}(10^7)$ higher and stem from $e^+e^-\to \Upsilon(4S)\to B\bar{B}$ and $e^+e^-\to q\bar q$ events, where $q$ indicates a $u,d,s$~or~$c$ quark. We therefore employ a $B$-tagging technique where one $B$ meson is fully reconstructed in a hadronic decay mode, and the signal is searched for among the remaining particles. We divide $\tau^+\tau^-$ final states into five mutually exclusive categories with similar sensitivities to signal. 
Backgrounds are suppressed using boosted decision tree (BDT) classifiers, whose output serves as the discriminating observable in a simultaneous fit to the Belle and Belle~II data. Background estimates are validated by comparing data and simulation in multiple background-enriched control channels.
To avoid experimenter bias, all selections are optimised on simulation and finalised before examining the signal region. 


\section{Detectors and samples} \label{sec:det&dat}

The Belle~II experiment is located at the SuperKEKB accelerator, which collides 7~GeV electrons and 4~GeV positrons at centre-of-mass (CM) energies near the $\Upsilon(4S)$
resonance~\cite{Akai:2018mbz,Ohnishi:2013fma}. The Belle~II detector~\cite{Abe:2010gxa} has a cylindrical geometry and includes a two-layer silicon-pixel detector~(PXD) surrounded by a four-layer double-sided silicon-strip detector~(SVD)~\cite{Belle-IISVD:2022upf} and a 56-layer central drift chamber~(CDC). Spatial hit information from these detectors is used to reconstruct the trajectories (tracks) of charged particles. Only one-sixth of the second layer of the PXD was installed at the time the data analysed here were recorded. The symmetry axis of these detectors, defined as the $z$ axis, nearly coincides with the electron beam direction; the transverse plane is the plane perpendicular to this axis. The detector is divided into forward endcap, barrel, and backward endcap regions according to the polar angle $\theta$ with respect to the $z$ axis, corresponding approximately to $\theta<32^\circ$, $32<\theta<130^\circ$, and $\theta>130^\circ$, respectively.
Surrounding the CDC, which also provides measurements of the specific ionisation energy loss by charged particles, are a time-of-propagation counter~\cite{Atmacan:2025jmh}, which measures the arrival time and detected position of Cherenkov photons propagating through quartz bar radiators in the barrel region, and an aerogel-based ring-imaging Cherenkov counter in the forward region. These detectors provide charged-particle identification information. They are surrounded by an electromagnetic calorimeter~(ECL) based on CsI(Tl) crystals,  which provides energy and timing measurements of electromagnetic showers for electron and photon identification. Outside the ECL is a superconducting solenoid magnet that provides a 1.5~T magnetic field parallel to the $z$ axis. Its flux return is instrumented with resistive-plate chambers and plastic scintillator modules (KLM) to detect muons, $K^0_{\rm L}$ mesons, and neutrons.
\\ \indent
The Belle detector~\cite{Abashian2002117, Brodzicka:2012jm} was located at the KEKB accelerator that collided 8~GeV electrons and 3.5~GeV positrons at CM energies near the $\Upsilon(4S)$ resonance~\cite{Kurokawa:2001nw,Abe:2013kxa}, and had an overall structure similar to Belle~II.
Vertex reconstruction and tracking were performed with the SVD and CDC detectors, which were different devices from those used in Belle~II.
For particle identification, the Belle detector employed aerogel threshold Cherenkov counters and a barrel-shaped array of time-of-flight scintillation counters, and the KLM relied exclusively on resistive-plate chambers. 
\\\indent For both detectors, the nearly $4\pi$ acceptance is essential in searches for decays involving neutrinos, as it allows the missing energy and momentum carried by undetected particles to be inferred from known collision kinematic properties and conservation laws.
\\ \indent
The collision data at the $\Upsilon(4S)$ resonance comprise the full Belle dataset, recorded during 1999--2010 and corresponding to an integrated luminosity of $711~\mathrm{fb}^{-1}$, together with Belle~II data recorded during 2019--2022, corresponding to an integrated luminosity of $365~\mathrm{fb}^{-1}$.
\\ \indent
Monte Carlo simulated events are used to model both signal and background processes and to optimise the event selection and analysis procedure. We use the \texttt{KKMC} generator~\cite{Jadach:1999vf} for $e^+e^- \to q\bar q$ events and the \texttt{EvtGen}~\cite{Lange:2001uf} software package for $B\bar B$ events. Event generation is interfaced with \texttt{PYTHIA8} (\texttt{PYTHIA6} for Belle)~\cite{uffi,Sjostrand:2014zea} for fragmentation, hadronisation, and decay modelling, and with \texttt{PHOTOS}~\cite{Barberio:1993qi} to account for final-state radiation. The simulated background samples are two and four times the size of the $\Upsilon(4S)$ datasets collected at Belle and Belle~II, respectively. In Belle, 
separate samples of $b\to u\ell^-\bar{\nu}_\ell\;(\ell=e,\mu)$ and other, rarer $b\to (s,d,u)$ processes are produced with an effective luminosity equal to 20 and 50 times the Belle $\Upsilon(4S)$ dataset, respectively. 
The simulation includes beam-induced background data overlay~\cite{LIPTAK2022167168} and a full detector response via the \texttt{Geant4} (\texttt{Geant3} for Belle) software package~\cite{Brun:1994aa,Agostinelli:2002hh}. The \texttt{EvtGen} model from Ref.~\cite{btosllball} is used to produce a set of signal events corresponding to a luminosity (assuming SM branching fractions) of roughly $2\times10^6$ times the experimental data sample. All collision data and simulations are processed with the Belle~II analysis software~\cite{Kuhr:2018lps}. Belle data samples are converted from the Belle analysis software~\cite{KATAYAMA199822} format into the Belle~II format for compatibility~\cite{Gelb:2018agf}.

\section{Event selection}\label{sec:evtsel}

A trigger selects events online based on the total energy and charged- and neutral-particle multiplicities to preferentially retain hadronic events, with full efficiency for signal events.

We select events with $\Upsilon(4S)$ candidates decaying into a pair of neutral $B$ mesons. 
One of the neutral $B$ mesons, referred to as the tag-side $\bar{B}^0$ meson (\bobtag), is formed using the full event interpretation (FEI) algorithm~\cite{keck}. The FEI reconstructs exclusive hadronic $B$-meson decay chains from final-state particles through intermediate baryon and meson candidates and assigns each $\bobtag$ candidate a classifier score $\mathcal{P}_\mathrm{FEI}$. The signal-side decay $B^0_{\rm sig}\to \KS \tau^+\tau^-$ is then selected from the particles not used in the \bobtag reconstruction. This tagging approach not only reduces the $q\bar{q}$ background, but also provides flavour and kinematic information on the \bosig, enabling discrimination against $B\bar{B}$ backgrounds. In $e^+e^-$ collisions with near-threshold $B\bar{B}$ production, the kinematic properties of the initial system are known, allowing the \bosig four-momentum to be constrained using the reconstructed \bobtag and energy conservation.


After imposing a minimal threshold on $\mathcal{P}_\mathrm{FEI}$ to increase the probability of a correct reconstruction,
we require the beam-constrained mass to satisfy
\begin{equation}\label{eq:mbc}
M_{\rm bc}=\sqrt{s/(4c^4)-|\vec{p}^{\,*}_{\bobtag}|^2/c^2} > 5.27~{\rm GeV}/c^2
\end{equation}
and the energy difference to satisfy
\begin{equation}\label{eq:deltae}
-0.15<\Delta E = E^*_{\bobtag} - \sqrt{s}/2 < 0.10~\text{GeV},
\end{equation}
where $\sqrt{s}$ is the total energy of the $e^+e^-$ system. Here and throughout this paper, $P_A$, $E_A$, $\vec{p}_A$, and $M_A$ denote the reconstructed four-momentum, energy, three-momentum, and invariant mass of particle $A$, respectively, with $P_A=(E_A/c,\vec{p}_A)$ and $M_A^2c^4=E_A^2-\vec{p}_A^{\,2}c^2$. For known masses we use the notation $m_A$. An asterisk indicates that the quantity is evaluated in the CM frame.
The selections in Eqs.~\ref{eq:mbc} and~\ref{eq:deltae} reduce combinatorial background and misreconstructed candidates with incorrectly estimated $\vec{p}^{\,*}_{\bobtag}$ and $E^*_{\bobtag}$, respectively.

In each event, the candidate with the highest $\mathcal{P}_\mathrm{FEI}$ is retained among the $\bobtag$ candidates that satisfy the requirements listed above. For simulated signal events, this selection achieves a fraction of correctly reconstructed $\bobtag$ mesons of approximately $70\%$, at an efficiency of around $0.5\%$.

On the signal side, the $\KS$ candidate is reconstructed via its decay to $\pi^+\pi^-$ due to its larger branching fraction, lower background, and better momentum resolution compared to other decay modes. Pairs of oppositely charged particles assumed to be pions are required to 
satisfy $450<M(\pi^+\pi^-)<550$~MeV/$c^2$.
We apply a selection in bins of $\KS$ momentum optimised using the following observables: the distance of closest approach to the interaction point in the transverse plane of each pion track;
the projection onto the transverse plane of the angle between the momentum of the $\KS$ candidate and the vector pointing from the interaction point to the $\pi^+\pi^-$ vertex; the longitudinal separation of the two pion tracks at their reconstructed vertex; and the transverse flight length of the $\KS$ candidate~\cite{gBKS}. This selection retains 80\% (87\%) of true $\KS$ candidates
in Belle (Belle~II) data and yields a $\KS$ misreconstruction rate of slightly below $2\%$ as estimated from signal simulation. 
\\\indent The $\tau^+\tau^-$ invariant mass squared, $q^2=(P_{\tau^+}+P_{\tau^-})^2$, provides additional discrimination against background and is derived from the four-momenta of the other reconstructed particles in the event and the known collision kinematic properties as 
\begin{equation}
    q^2_{\rm rec}= m_{B^0}^2 + M_{\KS }^2-2\left[\sqrt{s}E_{\KS }^*/(2c^4)+\vec{p}_{\bobtag}^{\,*}\cdot \vec{p}_{\KS }^{\,*}/c^2\right],
\end{equation} 
where $\vec{p}_{\bobtag}^{\,*}=-\vec{p}_{B^0_{\rm sig}}^{\,*}$, as in $e^+e^-\to\Upsilon(4S)\to B\bar{B}$ events the two $B$ mesons are back-to-back in the CM frame.
The $q^2_{\rm rec}$ resolution is improved by replacing $E_{\bobtag}^*$ and $M_{\bobtag}$ with $\sqrt{s}/2$ and the known $B^0$ mass $m_{B^0}$~\cite{ParticleDataGroup:2024cfk}, respectively. 
We require $q^2_{\rm rec} > 12~\mathrm{GeV}^2/c^4$, noting that the kinematic threshold is $q^2=4m_\tau^2\approx12.6~\mathrm{GeV}^2/c^4$ for signal decays, whereas background processes can populate smaller $q^2$ values. 
\\\indent The $\tau$ candidates are reconstructed in the four principal decay modes characterised by a single charged particle: two leptonic ($\tau^- \to \mu^-  \bar\nu_\mu \nu_\tau$ and $\tau^- \to e^-  \bar\nu_e\nu_\tau$) and two hadronic ($\tau^- \to \pi^- \nu_\tau$ and $\tau^- \to \rho^- \nu_\tau$, with $\rho^-\to\pi^-\pi^0$). Other modes with a single charged particle ($\tau^-\to K^-n\pi^0\nu_\tau,\pi^- m\pi^0\nu_\tau$, with multiplicities $n\ge0,m\ge1$) are not explicitly reconstructed but can enter the selected sample. The signal simulation is used for the selection optimisation and includes a mixture of $\tau$ decay modes with their branching fractions corrected according to their known values~\cite{ParticleDataGroup:2024cfk}.
The charged particles considered as $\tau$-decay products
are selected from tracks that are not associated with the \bobtag or $\KS $ candidates. These charged-particle candidates must have momenta above $50$~MeV$/c$ and originate from the interaction point.
\\\indent Leptons are identified using discriminating observables $\mathcal{P}^\ell$ that combine information from multiple subdetectors. In Belle, electron identification relies on the ECL, CDC, and aerogel threshold Cherenkov counters~\cite{belle_eid}, while muons are identified with the KLM~\cite{belle_muid}; in Belle~II, all subdetectors except the PXD and SVD contribute~\cite{lidbelleii}. The discriminating observables are built from likelihood ratios, except for electrons in Belle~II data, where a BDT classifier is used. We require $\mathcal{P}^{\ell} > 0.9$, $|\vec{p}_\ell| > 200~\mathrm{MeV}/c$ and polar-angle ranges $17^\circ$--$150^\circ$ ($18^\circ$--$151^\circ$) for muons (electrons) in Belle data, and $23^\circ$--$150^\circ$ ($13^\circ$--$155^\circ$) in Belle~II data. The resulting efficiencies are 44\% (77\%) for muons (electrons) in Belle data and 81\% (67\%) in Belle~II data, while the probabilities for other charged particles to be misidentified as muons (electrons) are 6\% (below 1\%) in Belle data and 19\% (below 1\%) in Belle~II data, as estimated from simulated signal events.  
\\ \indent Neutral pions are reconstructed from pairs of photons. Photon candidates are localised energy deposits (clusters) in the ECL that are not associated with charged particles and that satisfy the following minimum-energy requirements: 120~MeV, 30~MeV, and 80~MeV in the forward, barrel, and backward regions of the calorimeter, respectively. We restrict the polar angle of photon candidates to the range $17^\circ$--$150^\circ$ and apply quality requirements on the number of crystals in the cluster and on the energy distribution among the crystals within the cluster. 
Photon pairs are combined to form $\pi^0$ candidates if their invariant mass is in the range $(121, 142)$~MeV$/c^2$, and if the diphoton opening angle projected onto the transverse plane is smaller than 1.0 rad and the three-dimensional opening angle is smaller than 0.9 rad. We require the $\pi^0$ to have a reconstructed momentum in the range $(0.2,3.0)$~GeV$/c$. We reconstruct up to two $\pi^0$ candidates per event to account for signal topologies where both $\tau$ leptons decay via $\tau^\pm \to \rho^\pm\overset{\scriptscriptstyle(-)}{\nu}_\tau$, with $\rho^\pm\to\pi^\pm\pi^0$.
\\ \indent Each charged $\tau^\pm$-decay-product candidate $t^\pm$ is tested against particle-identification criteria. A candidate is first required to satisfy the muon selection; if it fails, the electron selection is applied. Tracks fulfilling either criterion are classified as leptons ($\ell$); otherwise, they are classified as hadrons ($h$). If such a hadron, together with the associated $\pi^0$, yields an invariant mass $M(\pi^\pm\pi^0)$ in the range $(480,1070)$ MeV/$c^2$, the charged decay product 
is considered to be a $\rho^\pm$ meson.
\\ \indent After selecting the \bobtag, $\KS$, and the two $\tau$-decay-product candidates with the appropriate charge combination, we require that no additional charged particles originating from the interaction point are reconstructed in the event. This selection suppresses backgrounds arising from combinations of incorrectly reconstructed $\bobtag$ and $B^0_{\rm sig}$ mesons, and simultaneously reduces the occurrence of signal events with $\tau$ decays into more than one charged particle. Additional ECL neutral clusters, not associated with the $\bobtag$ or the $B^0_{\rm sig}$ candidates, are permitted and used at a later stage for background suppression. Such clusters can originate from beam backgrounds, charged-hadron-induced backgrounds, or $B$-meson misreconstruction. 
\\ \indent 
The flavour of the $\bobtag$, obtained from exclusive reconstruction in a hadronic decay mode, provides further discrimination against background. 
While the signal final state is common to $B^0$ and $\bar{B}^0$ decays, the dominant $B\bar{B}$ backgrounds are not flavour symmetric since they arise through CKM-favoured $\bar{b} \to \bar c \to \bar s$ decay chains. The correlation of the tag-side $B$ flavour with the charge of the lepton candidate, which is valid up to $B^0$--$\bar B^0$ mixing effects,  
provides separation between
$\bar{b}\to \bar{c}\, \ell^+\bar\nu_\ell$ events and $\bar{c}\to \bar{s}\, \ell^-\nu_\ell$ events, improving background suppression for ${B}^0_{\rm sig}\to \KS \ell h$ signal candidates. We therefore use five mutually exclusive final-state categories: $\ell\ell$, $\rho\ell$, $\ell^- h^+$, $\ell^+ h^-$, and no-$\ell$.
At this stage, the mean candidate multiplicities in simulated signal events are below 1.03 for the $\ell\ell$, $\ell^- h^+$, and $\ell^+ h^-$ categories, while for the $\rho\ell$ and no-$\ell$, they are in the range 1.1--1.3. In cases with multiple candidates in a given category, we choose a candidate at random.
\\\indent The signal efficiencies and relative populations of the categories are evaluated using inclusive and signal simulations processed with the same reconstruction and selection as the data. Differences between data and simulation in tracking, lepton identification and misidentification, and $\pi^0$ reconstruction, which can cause migrations among the $\ell$, $h$, and $\rho$ final states, are corrected using calibration and control samples, as described in Section~\ref{sec:sys}.

\section{Background characterisation and suppression} \label{sec:bg}

The background composition varies depending on the $\tau^+\tau^-$ final state.
The dominant backgrounds are $B\to D(\to X_{\rm SL}) \ell \nu_\ell$ decays for the $\ell\ell$ category, $B\to D(\to X_{\rm HAD}) \ell \nu_\ell$ decays for both the $\ell^+ h^-$ and $\rho\ell$ categories,  $B\to D(\to X_{\rm SL}) X_{\rm HAD}$ decays for the $\ell^- h^+$ category, and $B\to D(\to X_{\rm HAD}) X_{\rm HAD}$ decays for the no-$\ell$ category, where $X$ denotes an arbitrary final state, and the subscript ``SL" (``HAD") indicates the presence (absence) of a charged lepton within $X$. 

In order to suppress the backgrounds above, we train a separate BDT classifier~\cite{fbdt} for each of the five categories for both Belle and Belle~II. The input observables are selected based on their discriminating power and the quality of their data-simulation agreement, which is confirmed in several signal-depleted samples (details are given in Section~\ref{subsec:control}). 
The input observables comprise the following quantities related to the signal side: the invariant mass of the \KS and the $t^-$, $M(\KS t^-)$, to characterise backgrounds where the two particles have the same origin (mostly a $D^-$ meson); the invariant mass of the \KS and the $\pi^0$ used for the $\rho$ candidates, $M(\KS \pi^0)$, for backgrounds with $D^0$ and $K^{*0}$ mesons; and the invariant mass of the $\pi^\pm\pi^0$ system, $M(\pi^\pm\pi^0)$, to reduce the combinatorial backgrounds. The $M(\pi^\pm\pi^0)$ observable is used only for the $\rho\ell$ and no-$\ell$ categories. We veto candidates with $M(\KS t^-)$ in the $(1.8,1.9)$~GeV/$c^2$ window whenever $t^-$ is a hadron to suppress the background from hadronic $D^-$ decays, and use the vetoed sample for validation in data. Another signal-side-related feature is the momentum $|\vec{p}^{\,*}_{t^+}|$ of the $t^+$ candidate, useful for reducing $B \to X\ell\nu_\ell$ backgrounds since the lepton from semileptonic $B$ decays typically has higher momentum than any signal product. Observables related to both the signal and tag sides are the following: $q^2_{\rm rec}$, where backgrounds tend to populate higher values; the cosine of the angle between the thrust axis 
of the $\bobtag$ and the thrust axis of everything else in the event~\cite{Belle:2012yvr}, to separate the more isotropic \BBbar events from the jet-like \qqbar events; the difference in the longitudinal distance between the $\bobtag$ and $B^0_{\rm sig}$ vertices, which also provides separation between \BBbar and \qqbar events due to the long lifetime of $B$ mesons. The $B^0_{\rm sig}$ vertex is determined from the intersection of the \KS flight trajectory with the $t^-$ track. 
Other observables are designed to capture the detector response not associated with the reconstruction of either $\bobtag$ or $B^0_{\rm sig}$, collectively referred to as the \textit{rest of the event} (ROE). A crucial one is the energy sum of the ROE deposits in the ECL ($E_{\rm ECL}$). After a signal event has been correctly reconstructed, no residual calorimeter activity should be present, whereas misreconstructed background events have larger extra ECL activity. Another useful observable is the sum of the missing energy and momentum in the event, $U={E}^\prime+c|\vec{{p}}^{\,\prime}|$, with ${P}'= ({E}'/c,\vec{p}^{\,\prime}) = P_{e^+e^-}-P_{B^0_{\rm sig}}-P_{\bobtag}-P_{\rm ROE}$, as larger missing momentum and energy are expected for signal events given the presence of up to four neutrinos. For the ROE-related ECL clusters, we require their energies to exceed 100 MeV, 50 MeV, and 150 MeV in the forward, barrel, and backward regions, respectively. We also impose a cluster isolation criterion with respect to the tracks of charged particles. For each cluster, we compute the distances along the inner surface of the calorimeter between the centre of its highest-energy crystal and the intersection points of tracks extrapolated from the drift chamber; the minimum distance must exceed 20~cm to reduce the contamination of neutral cluster candidates by ECL activity associated with charged hadrons.  
Finally, for the same reasons discussed for $U$, we consider the missing mass squared, defined as $E^2_{\rm miss}/c^4-|\vec{p}_{\rm miss}|^2/c^2$ with  $p_{\rm miss} = (E_{\rm miss}/c, \vec{p}_{\rm miss}) = P_{e^+e^-}-P_{B^0_{\rm sig}}-P_{\bobtag}$. 
For all the classifiers, $E_{\rm ECL}$ is among the two most-discriminating features. The other is $U$, except for the $\ell^+ h^-$ ($\ell\ell$) categories, where $|\vec{p}^{\,*}_{t^+}|$ (missing mass squared) has the highest importance.    
\\ \indent The BDT output score $\mathcal{O}$ is transformed into an observable $\mathcal{O}^\prime$ that is uniformly distributed for signal events using the probability integral transform~\cite{mutransf}. The transformed observable $\mathcal{O}^\prime$ is used in the final fit. The common requirement $\mathcal{O}^\prime > 0.3$  defines the fit region and retains 70\% of signal events while rejecting 80\%--95\% of background, depending on the category. The statistical gain obtained by extending the fit to lower $\mathcal{O}^\prime$ values does not compensate for the increase in systematic uncertainties related to $B$-decay modelling (discussed in Section~\ref{subsec:extinput}) in simulation arising from the larger retained background. The minimal requirement on the fit observable is not explicitly optimised and set to be the same for all categories for simplicity, with a contiguous region reserved for modelling validation in data. 
\\\indent Within the fit region, the residual background is dominated by \BBbar events, with \qqbar contributions accounting for at most 30\% of the total. The \BBbar background arises primarily from hadronic decays of the \btag combined with semileptonic decays of the \bsig. In particular, $\bsig\to D^{(*)}\ell\nu_\ell$ decays constitute 76\%, 44\%, 57\%, 35\%, and 21\% (67\%, 42\%, 54\%, 31\%, and 19\%) of the \BBbar background in the $\ell\ell$, $\ell^- h^+$, $\ell^+ h^-$, $\rho\ell$, and no-$\ell$ categories for Belle (Belle~II), respectively. The next-most-significant contribution arises from $B\to D^{(*)}\tau\nu_\tau$ decays, which account for about 10\% of the signal-side $B\bar B$ background in each category. The $B\to D^{(**)}\ell\nu_\ell$ and $B\to D^{(**)}\tau\nu_\tau$ decays contribute to less than 10\% of backgrounds, except for the $\rho\ell$ category, where this fraction ranges up to about 21\%. For the no-$\ell$ category, the backgrounds are mostly composed of fully hadronic decay chains with up to 6\% contribution of hadronic $B$ decays involving $D^{**}$. Our selection does not exclude the contribution from $B^0 \to \psi(2S)(\to \tau^+\tau^-)\KS$ decays, since its SM rate is below the experimental sensitivity and yields less than a single candidate.

\section{Signal extraction} \label{sec:fit}
The signal extraction is performed through a binned maximum likelihood fit to the modified BDT output observable $\mathcal{O}^\prime$ using templates constructed from simulation~\cite{caby,pyhf_joss}.

In each $\tau^+\tau^-$ category (denoted as $C$), three templates are considered: signal events, $B\bar{B}$ background, and $q\bar{q}$ background. The $B\bar{B}$ template includes both $B^0$ and $B^+$ events and both correctly and incorrectly reconstructed tag-side $B$-mesons.
All templates are weighted to correct for differences between data and simulation in reconstruction efficiency and normalisations as discussed in Section~\ref{sec:sys}; the average weights for the three templates are 0.77, 0.79, 0.81 (0.73, 0.64, 0.74) for Belle (Belle~II). Simulated templates are further scaled to the integrated luminosity of the~data.

The choice of binning for $\mathcal{O}^\prime$ is optimised per category using signal efficiency quantiles, with additional constraints to ensure sufficient event counts per bin. We use 7 bins for the $\ell\ell$ and $\rho \ell$ categories, 14 bins for $\ell^- h^+$ and $\ell^+ h^-$ categories, and 28 bins for the no-$\ell$ categories in both the Belle and Belle~II datasets.


The likelihood function $\mathcal{L}$ is a product of Poisson probability density functions $\mathcal{P}$ and combines the information from all bins of signal and background samples,
\begin{multline} \label{eq:likelihood}
\mathcal{L}(\mathcal{B}(B^0\to \KS \tau^+\tau^-)\mid \vec{\chi}) = \\
= \prod\limits_{\substack{C \in \\ \text{categories}}}
\prod\limits_{\substack{b \in \\ \text{bins}_C}}
{\mathcal{P}}(n_{C,b}\mid \nu_{C,b}(\mathcal{B}(B^0\to \KS \tau^+\tau^-),\vec{\chi})) 
\times \prod_{\chi \in \vec{\chi}} {\mathcal{G}}(a_\chi\mid \chi),
\end{multline}
where $n_{C,b}$ is the number of observed events in bin $b$ of category $C$, and the corresponding expected value is denoted by $\nu_{C,b}$.
The systematic uncertainties discussed in the next section are embedded into the likelihood through a set of nuisance parameters $\vec{\chi}$, which are event count modifiers constrained by normal distributions $\mathcal{G}$ according to auxiliary information $a_{\chi}$. The expected number of events in each bin is
\begin{multline} \label{eq:expyields}
\nu_{C,b}(\mathcal{B}({B^0\to \KS \tau^+\tau^-}),\vec{\chi}) = \\
 = 2 N_{\Upsilon(4S)} f_{00}  \mathcal{B}(B^0\to \KS \tau^+\tau^-) {\varepsilon_{C,b}(\vec{\chi})}+  \nu^{B\bar{B}}_{C,b}(\vec{\chi}) + \nu^{q\bar{q}}_{C,b}(\vec{\chi}),
\end{multline}
where $N_{\Upsilon(4S)}$ is the number of produced $\Upsilon(4S)$ mesons and $f_{00}$ is the $\Upsilon(4S) \to B^0\bar{B}^0$ branching fraction. The branching fraction $\mathcal{B}({B^0\to \KS \tau^+\tau^-})$, free in the fit, sets the normalisation of all signal templates and is extracted from a simultaneous fit across all categories. The signal efficiency $\varepsilon_{C,b}$ includes all the simulated $\tau$-decay modes with branching fractions fixed to their known values~\cite{ParticleDataGroup:2024cfk}. The efficiency for every category is computed as a ratio of number of signal events in the selected sample to the total number of generated events. The $\nu^{B\bar{B}}_{C,b}$ and $\nu^{q\bar{q}}_{C,b}$ denote the yields of the $B\bar{B}$ and $q\bar{q}$ templates, respectively. Studies with an ensemble of simplified experiments show that the branching-fraction estimator is unbiased.

Under the condition that no significant signal is observed in data, we set an upper limit on the 
$B^0\to \KS \tau^+\tau^-$ branching fraction using the CL$_\text{s}$ technique~\cite{CLs1,CLs2,junk_cls} with the asymptotic approximation of the profile likelihood ratio test statistic.
The five fit categories exhibit different sensitivities to $B^0\to \KS \tau^+\tau^-$ decays. Categories with at least one leptonic $\tau$ decay show superior performance, primarily due to reduced $q\bar{q}$ and hadronic $B$-decay backgrounds. In a fit model that includes only statistical uncertainties, the no-$\ell$ category provides the highest sensitivity owing to its larger signal efficiency, while the remaining categories exhibit comparable performance. However, when the full model with systematic uncertainties is considered, categories with larger hadronic backgrounds are more strongly impacted by modelling uncertainties. Consequently, the purely leptonic category provides the highest sensitivity, while the remaining categories still make non-negligible contributions and improve the final result. In each category, the Belle data contributes more to the overall sensitivity than Belle~II due to the larger size of the dataset. Table~\ref{table-with-yields} shows the numbers of background events expected ($\nu^{\rm bg}$), signal efficiencies ($\varepsilon^{\rm sig}$) and expected upper limits at the 90\% confidence level. The expected upper limits are evaluated on background-only simulated samples with the nominal values of the nuisance parameters. 

\begin{table}[ht]
\centering
\caption{Expected background yields, signal efficiencies, and expected upper limits on $\mathcal{B}(B^0\to \KS \tau^+\tau^-)$ at the 90\% confidence level per category.} \label{table-with-yields}
\setlength{\tabcolsep}{5pt} 
\begin{tabular}{lccccc|ccccc}

\hline
 & \multicolumn{5}{|c|}{\textbf{Belle}} & \multicolumn{5}{c}{\textbf{Belle~II}} \\
 &\multicolumn{1}{|c}{$\ell\ell$}  & $\ell^-h^+$ & $\ell^+h^-$ &$\rho\ell$ &no-$\ell$ &$\ell\ell$ &$\ell^-h^+$ &$\ell^+h^-$ &$\rho\ell$ &no-$\ell$ \\
\hline
$\nu^{\rm bg}$  &\multicolumn{1}{|c}{442} &1459 &2109 &632 &16186 & 618 &1147 &1317 &925 & 6745 \\
$\varepsilon^{\rm sig}\ (10^{-5})$  &\multicolumn{1}{|c}{3.4}  &8.0 &8.2 &3.1 & 28.2& 6.8 & 10.8 &10.4 & 7.4 & 30.3 \\

Limit $(10^{-4})$ &\multicolumn{1}{|c}{16.9}  &19.2 &18.9 &19.7 &24.1 &20.6  &26.3  &24.6 &23.4 &25.2 \\

\hline
\end{tabular}
\end{table}


\section{Systematic uncertainties and control samples} \label{sec:sys}

Systematic uncertainties enter the fit as nuisance parameters that affect either the template normalisation, their shapes, or both. 
The uncertainties related to the limited simulated sample size are included as shape systematics that are uncorrelated across bins, following Ref.~\cite{hifa}. Additionally, we consider the uncertainties on the external inputs, those associated with the auxiliary measurements for efficiency calibrations, and the uncertainties on the measurements performed in the data sidebands.
\\\indent The sources of systematic uncertainty and their contributions to the total uncertainty on $\mathcal{B}({B^0\to \KS \tau^+\tau^-})$ are summarised in Table~\ref{systable}. For each source, the contribution to the total uncertainty is computed following Ref.~\cite{rankPinto}. 



\begin{table}[h!]
\centering
\caption{Sources of systematic uncertainties, their treatment in the fit
and their contribution $\sigma_\mathcal{B}$ to the uncertainty on $\mathcal{B}({B^0\to \KS \tau^+\tau^-})$. The uncertainty type can be ``Normalisation", corresponding to a global
normalisation factor common to all bins, ``Correlated shape", for fully correlated bin-dependent uncertainties, and ``Uncorrelated shape", for bin-wise independent uncertainties. Each source is described by one or more nuisance parameters.
\label{systable}}

\begin{tabular}{l|l|c}
\hline
\textbf{Source}&\makecell[l]{\textbf{Uncertainty type,}\\\textbf{number of parameters}}&$\mathbf{ \sigma_\mathcal{B} (10^{-4})}$\\
\hline
Leading $B$-decay branching fractions & Correlated shape, 140 & \textcolor{black}{2.0}\\
Simulation sample size & Uncorrelated shape, 140 & \textcolor{black}{1.8}\\
$B$-tagging efficiency & Correlated shape, 2 & \textcolor{black}{0.6} \\
Combinatorial \bobtag normalisation & Correlated shape, 10 & \textcolor{black}{0.5}\\
$D \to \KL$ modelling& Correlated shape, 2 & \textcolor{black}{0.5} \\
Lepton ID & Correlated shape, 6 & \textcolor{black}{0.3} \\
$f_{00}$& Correlated shape, 1 & \textcolor{black}{0.2} \\
Signal efficiency uncertainty& Normalisation, 2 & \textcolor{black}{0.2} \\
$\pi^0$ efficiency & Correlated shape, 2 & \textcolor{black}{0.2} \\
$q\bar{q}$ normalisation & Normalisation, 4 &\textcolor{black}{0.2} \\
Signal model & Correlated shape, 1 & \textcolor{black}{0.1} \\
$\KS $ efficiency & Correlated shape, 2 & \textcolor{black}{$<$0.1} \\
Number of $\Upsilon(4S)$ & Normalisation, 2 & \textcolor{black}{$<$0.1} \\
Luminosity & Normalisation, 2 &\textcolor{black}{$<$0.1} \\
Tracking efficiency & Correlated shape, 2 &\textcolor{black}{$<$0.1} \\
Leading $\tau$-decay branching fractions & Correlated shape, 10 & \textcolor{black}{$<$0.1} \\
$\chi_d$ & Correlated shape, 1 &\textcolor{black}{$<$0.1} \\
\hline
\textbf{Total} & \multicolumn{1}{c}{\textbf{\ }} &\multicolumn{1}{c}{\textbf{  \textcolor{black}{2.9}   }} \\ 
\hline
\end{tabular}
\end{table}

\subsection{External inputs} \label{subsec:extinput}
The total number of $\Upsilon(4S)$ mesons is determined to be $(772 \pm 11)\times10^6$ for the Belle and $(387 \pm 6)\times10^6$ for the Belle~II datasets. The respective relative uncertainties, 1.4\% and 1.6\%, are taken as normalisation systematic uncertainties on the signal templates.
The value of $f_{00}$ is $0.4861\pm0.0080$~\cite{f00hflav}, and the relative uncertainty is implemented as a shape uncertainty, correlated across bins for both the signal and the $B\bar{B}$ background templates.
The uncertainties on the recorded integrated luminosities, 1.4\% in Belle~\cite{Brodzicka:2012jm} data and $0.5\%$ in Belle~II~\cite{lumiB2} data, are implemented as normalisation uncertainties on the simulated background templates. The $0.6\%$ uncertainty on the integrated mixing rate $\chi_d$~\cite{ParticleDataGroup:2024cfk} is also included in the fit model as a shape systematic, which allows the relative fractions of $B^0\bar{B}^0$ and $B^0{B}^0/\bar{B}^0\bar{B}^0$ events to vary.
\\\indent As an alternative signal model, we use the SM predictions from Ref.~\cite{Straub:2018kue}, which relies on different form-factor parameterisations and Wilson-coefficient inputs~\cite{Gubernari:2023puw,Aebischer:2018iyb}. The resulting variation in the $q^{2}$ spectrum is treated as a correlated shape uncertainty across all bins of the signal template.
\\\indent The main background contributions for the $B^0_{\rm sig}$ candidates ($B \to D^{(*,**)} \ell \nu_\ell$, $B \to D^{(*,**)} \tau \nu_\tau$ and measured hadronic decays) are corrected for mismatches between the generated rates in the simulation and the latest experimental averages from Refs.~\cite{ParticleDataGroup:2024cfk,f00hflav}. The uncertainties on these averages, typically in the range of $2\%$--$6\%$, are also included in the fit. For $B$ decays that have not been measured experimentally or that give a small contribution to the simulated background (e.g. hadronic modes with neutrons, decays involving higher-resonance $D$ mesons, or $B \to D^{(*,**)} \ell \nu_\ell X$ and $B \to D^{(*,**)} \tau \nu_\tau X$ modes), a 100\% uncertainty on their abundance is considered.
The fraction of events whose simulated rates have been corrected for the \bosig candidate in $B\bar B$ background is about 70\%--90\% in categories with at least one reconstructed lepton, and about 50\% in the no-$\ell$ category. 

These corrections affect both the shapes and the normalisations of the $B\bar{B}$ templates. We perform studies with an ensemble of simplified experiments to capture the correlations among the systematic uncertainties related to the branching fractions of $B$-meson decays across the fitting categories and the Belle and Belle~II datasets. 
We use this ensemble to construct a joint covariance matrix, and find the 140 eigenvectors corresponding to correlated shape variations across the bins. Each variation is represented by a nuisance parameter.

We also consider the uncertainties on the $\tau$-decay branching fractions from the latest experimental averages~\cite{ParticleDataGroup:2024cfk} used to correct the simulation; for each of the ten largest modes, an independent nuisance parameter is used.
\\\indent Following Ref.~\cite{btoknunu}, the fraction of $D$-meson decays involving $\KL$ is assigned a 40\% normalisation uncertainty within the $B\bar{B}$ template. This uncertainty is treated as correlated across bins of the $B\bar{B}$ templates, independently for Belle and Belle~II data.

\subsection{Efficiency calibrations} \label{subsec:effsys}
The efficiency differences between data and simulation related to tracking, lepton identification and \KS and $\pi^0$ reconstruction are corrected based on auxiliary measurements. The uncertainties on the corrections are propagated as systematic uncertainties in the fit. 
\\\indent Corrections to the simulation and the associated uncertainties for the reconstruction of tracks of charged particles with momenta above 200 MeV$/c$ are evaluated in Belle using a low-background sample of $D^{*+} \to D^0(\to \KS \pi^+\pi^-) \pi^+$ decays. One pion from the $\KS$ decay serves as the probe track, while the kinematic properties of the decay are constrained using the remaining pion from the $\KS$, the two pions from the $D^0$, and the pion from the $D^{*+}$ decay~\cite{Bevan:2014iga}. In Belle~II, the efficiency is determined using $e^+e^- \to \tau^+\tau^-$ events with $\tau^+\to \ell^+\nu_\ell\bar{\nu}_\tau$ and $\tau^-\to \pi^-\pi^-\pi^+\nu_\tau$, where one pion from the hadronic $\tau$ decay is used as the probe~\cite{Bevan:2014iga}. In both Belle and Belle~II simulations, no efficiency corrections are needed and per-track systematic uncertainties of 0.35\% and 0.27\% are assigned, respectively. 
For tracks of charged particles with momenta below~200~MeV$/c$, we apply momentum-binned corrections with uncertainties of approximately 1\% obtained from $B^0 \to D^{*-}(\to \bar{D}^0\pi^-)\pi^+$ calibration samples, in which the pion from the $D^{*-}$ decay probes the low-momentum regime~\cite{Bevan:2014iga}. Such low-momentum charged particles constitute only about 10\% of the selected sample. All tracking uncertainties are implemented as correlated shape variations across all templates for Belle and Belle~II data via two independent nuisance parameters.
\\\indent Efficiencies for lepton identification and rates of pion-to-muon misidentification are corrected in simulated samples as functions of momentum and polar angle using weights derived from data control channels. The lepton identification efficiencies are determined using $e^+e^- \to \ell^+\ell^-\gamma$ (Belle~II only), $e^+e^- \to e^+e^- \ell^+\ell^-$, and $J/\psi \to \ell^+ \ell^-$ events. The $\pi \to \mu$ misidentification rates are obtained from $\KS \to \pi^+\pi^-$ decays and, for Belle~II, supplemented by $e^+e^- \to \tau^+(\to \pi^+\pi^+\pi^- \bar{\nu}_\tau)\tau^-$ events. In Belle (Belle~II), the mean systematic uncertainties are 2.6\% (1.9\%) for electron identification efficiency, 2.1\% (0.4\%) for muon identification efficiency, and 4.2\% (5.4\%) for incorrect identification of pions as muons. Due to their sub-percent contributions, the $\pi\to e$ components are not corrected, nor is a related uncertainty considered. Lepton identification uncertainties are implemented as shape uncertainties correlated across bins for all categories with leptons, and affect all templates.
\\\indent The differences in \KS vertex-reconstruction efficiency between data and simulation are measured with $D^{*+}\to D^0(\to \KS \pi^+\pi^-)\pi^+$ events as functions of kinematic observables. In Belle~II data they are also further binned in the \KS  vertex displacement. Corrections are applied to account for the observed discrepancies. The related systematic uncertainty is on average 0.6\% in Belle and 2.1\% in Belle~II signal simulations and is implemented as a shape uncertainty correlated across bins with one nuisance parameter for Belle and one for Belle~II.
\\\indent The \piz reconstruction efficiency is corrected using $\tau^-\to\pi^-\pi^0\nu_\tau$ samples for Belle data and $D^0\to K^-\pi^+\pi^0$ in Belle~II data. The differences observed between data and simulation are corrected for, and the mean systematic uncertainty is 2.2\% (4.1\%) in Belle (Belle II) data. The $\pi^0$ efficiency uncertainty is implemented as a shape uncertainty correlated across bins in the $\rho\ell$ and no-$\ell$ categories for all the templates.

\subsection{Measurements in control samples} \label{subsec:control}
Multiple validation channels and kinematic sidebands are employed to validate the modelling accuracy of the simulation. Samples collected at a CM energy about 60 MeV below the $\Upsilon(4S)$ resonance, where \BBbar production is kinematically forbidden, are used to correct the $q\bar{q}$ background component. The integrated luminosities for these samples are 90~fb$^{-1}$ and 42~fb$^{-1}$ for Belle and Belle II, respectively. Normalisation corrections are derived for simulated events in the fit region for the no-$\ell$ category and for the combined $\ell\ell$, $\ell^- h^+$, $\ell^+ h^-$, and $\rho\ell$ categories. In Belle (Belle II) the corresponding corrections are $0.67\pm0.03$ and $0.57\pm0.12$ ($0.71\pm0.05$ and $1.02\pm0.16$). The uncertainties on the corrections are of statistical nature and are taken as normalisation systematic uncertainties on the \qqbar templates. 
\\\indent A dedicated study of the hadronic $B$-tagging performance is carried out to correct the simulated signal and $B\bar{B}$ background yields, and to validate their modelling. 
The $B\bar{B}$ background events are divided into two categories, depending on whether the tag-side $B$ meson is correctly or incorrectly reconstructed. In the former case, the candidates exhibit a prominent peak in $M_{\rm bc}$.

The tagging efficiency of events with a properly reconstructed \bobtag is corrected for differences between data and simulation arising from the branching fractions and decay models used for the relevant hadronic $B$ decays. The correction is determined by comparing the number of events containing both a $\bobtag$ candidate and a low-background signal-side decay such as $B^0 \to X^-\ell^+\nu_\ell$ in Belle~II data and $B^0 \to D^-\pi^+$ in both Belle and Belle~II data. The correction factors, which depend on the $\bobtag$ decay mode, are on average 75.3\% (76.7\%), with uncertainties of 5.2\% (5.7\%), for the Belle (Belle~II) dataset. The signal and $B\bar{B}$ efficiencies are corrected in simulation using these calibration factors, and the associated uncertainties are implemented as shape variations correlated across bins separately for each dataset.


An incorrectly reconstructed \bobtag candidate is defined by the presence of one or more wrongly associated reconstructed particles, such as tracks or clusters originating from the other $B$ meson. The normalisation of this component in the $B\bar{B}$ template can differ between data and simulation because it receives contributions from a different mixture of $B$-decay modes. We therefore check the agreement of simulation with data in signal-depleted samples and assign a 50\% uncertainty, implemented in the fit as one shape uncertainty correlated across bins per category.
Two background control samples are used for this validation: the $\KS$ sideband and the $M_{\rm bc}$ sideband. The $\KS$ sideband is defined by selecting $\pi^+\pi^-$ candidates in the regions
$0.480<M(\pi^+\pi^-)<0.486~{\rm GeV}/c^2$ and $0.510<M(\pi^+\pi^-)<0.516~{\rm GeV}/c^2$,
while also inverting the $\KS$ quality criteria to increase the sample size. In this region the signal is absent, but the kinematic properties of the remaining backgrounds are similar to those in the fit region. The simulation shows good agreement with data in the BDT output and input features, validating the background composition and modelling.
The $M_{\rm bc}$ sideband, defined by
$5.24 < M_{\rm bc}<5.27~{\rm GeV}/c^2$,
is enriched in incorrectly reconstructed $\bobtag$ candidates and is used to verify the combinatorial component of the $B\bar{B}$ template. 
The normalisation factors derived in this sideband agree within statistical uncertainties with those measured in the \KS sideband except for the no-$\ell$ ($\ell^- h^+$) categories in Belle (Belle~II), motivating the 50\% systematic uncertainty on the normalisations.

The $M(\KS t^-)$-vetoed region is enriched in semileptonic $B^0\to D^{(*)-}\ell^+ \nu_\ell$ decays with $D^-\to \KS\pi^-$ or $\KS\rho^-$. 
We use this sample to validate the signal extraction procedure by performing fits to the $\mathcal{O}^\prime$ observable with data-to-simulation $D^{(*)-}$ yield ratios as parameters of interest. We obtain $1.09 \pm 0.11$ $(1.04 \pm 0.14)$ for these ratios, confirming good agreement between data and simulation. Such agreement is also observed in the BDT input observables. 
We evaluate the uncertainty associated with the $\mathcal{O}^\prime$ selection efficiency using the same control sample. Although the relevant BDT inputs are well modelled and provide strong discrimination between signal and background in the nominal signal selection, the $B^0\to D^{(*)-}\ell^+ \nu_\ell$ control sample is kinematically different from signal by construction. In particular, $M(\KS t^-)$ is restricted by the definition of the control region, while the missing mass squared peaks near zero for these semileptonic decays. To prevent these control-sample-specific kinematic features from dominating the validation, the values of these two observables are drawn in both data and simulation from the corresponding signal distributions before evaluating the BDT output. The resulting $\mathcal{O}^\prime$ distribution is approximately uniform, allowing the efficiency of the $\mathcal{O}^\prime > 0.3$ requirement to be compared between data and simulation. The obtained selection efficiencies agree within statistical uncertainties of about 10\%, which are propagated as multiplicative uncertainties to the Belle and Belle~II signal templates, with one nuisance parameter assigned to each dataset.

A fit to data in the signal-depleted region $0.15\leq\mathcal{O}^\prime \leq0.30$ is used to validate the background modelling and check that the fitted normalisations and shapes adequately reproduce the observed data. The agreement of simulation with data in BDT input features is confirmed in this background-enhanced region.

\section{Results} \label{sec:results}

\begin{figure}[!ht]
    \centering
    \begin{subfigure}{0.3\textwidth} 
        \includegraphics[width=\linewidth]{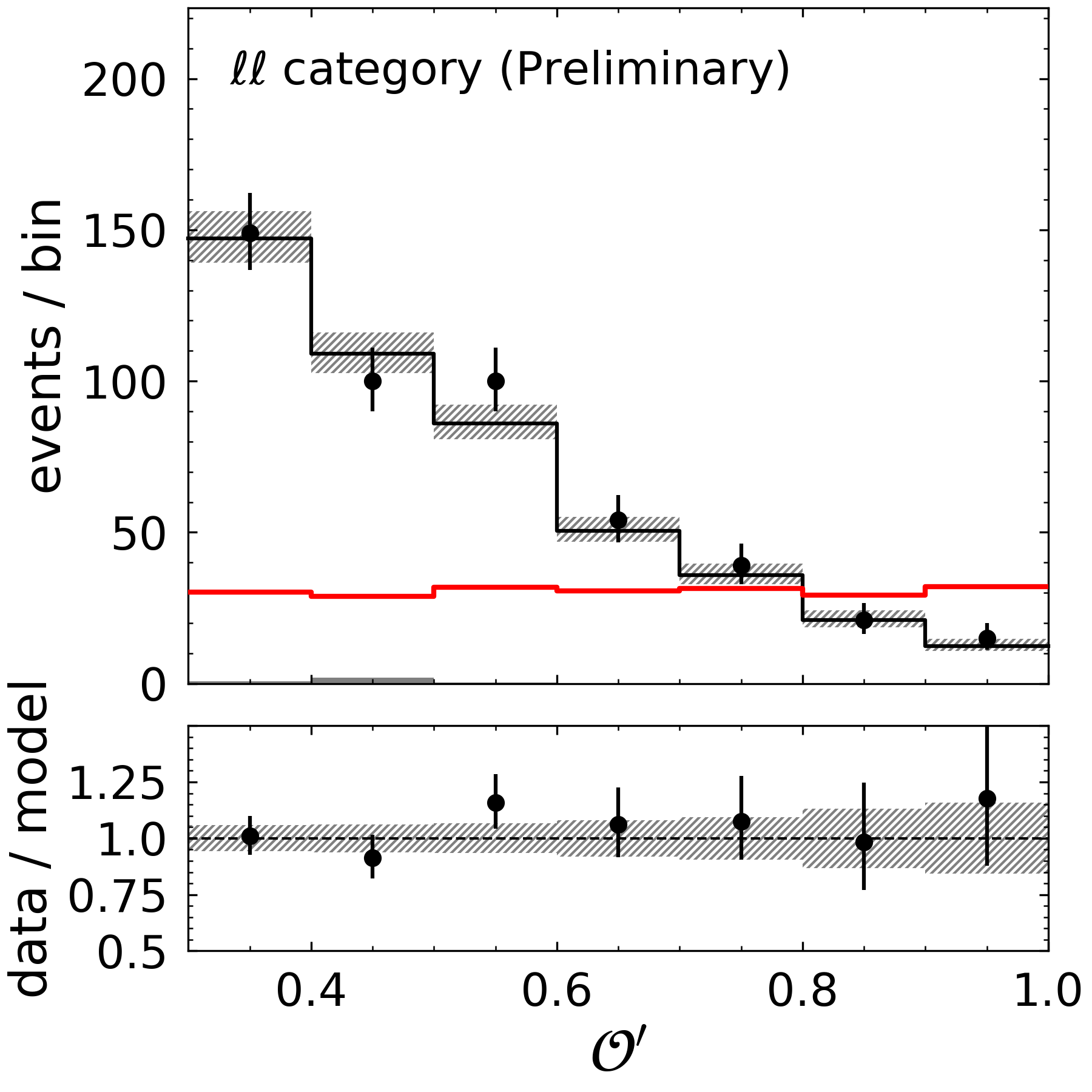}

    \end{subfigure}
    \begin{subfigure}{0.3\textwidth} 
        \includegraphics[width=\linewidth]{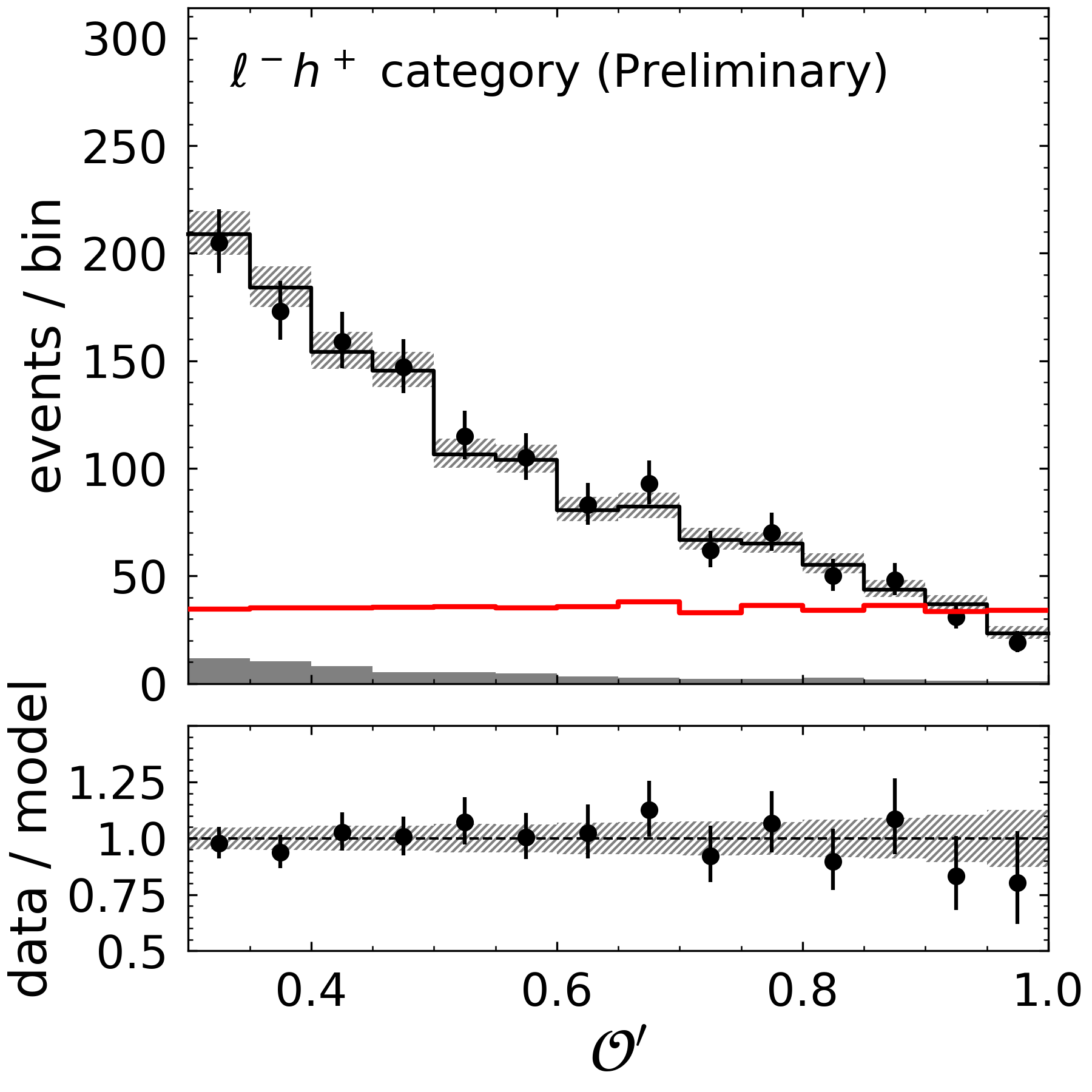}

    \end{subfigure}
    \begin{subfigure}{0.3\textwidth} 
        \includegraphics[width=\linewidth]{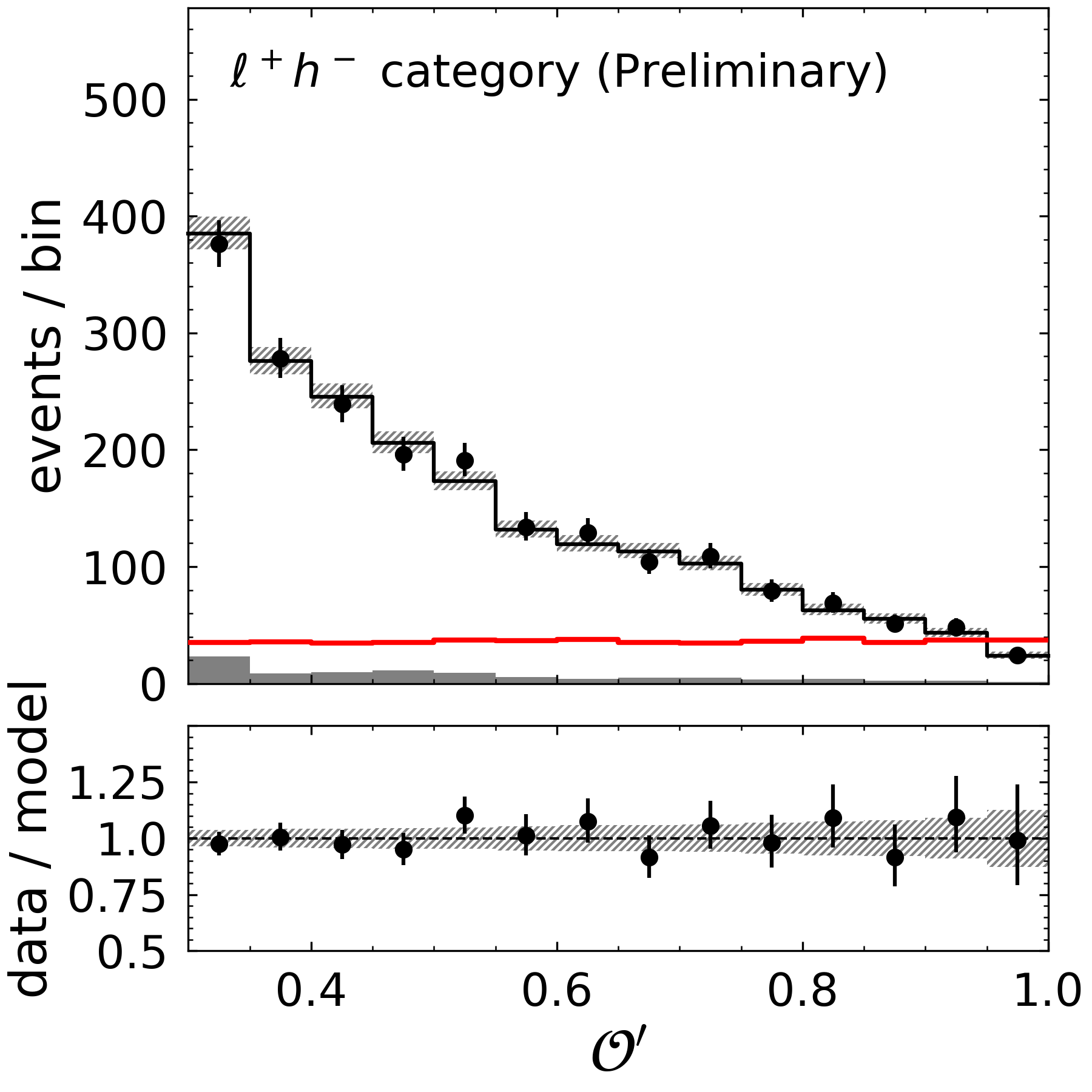}

    \end{subfigure}
    \begin{subfigure}{0.3\textwidth} 
        \includegraphics[width=\linewidth]{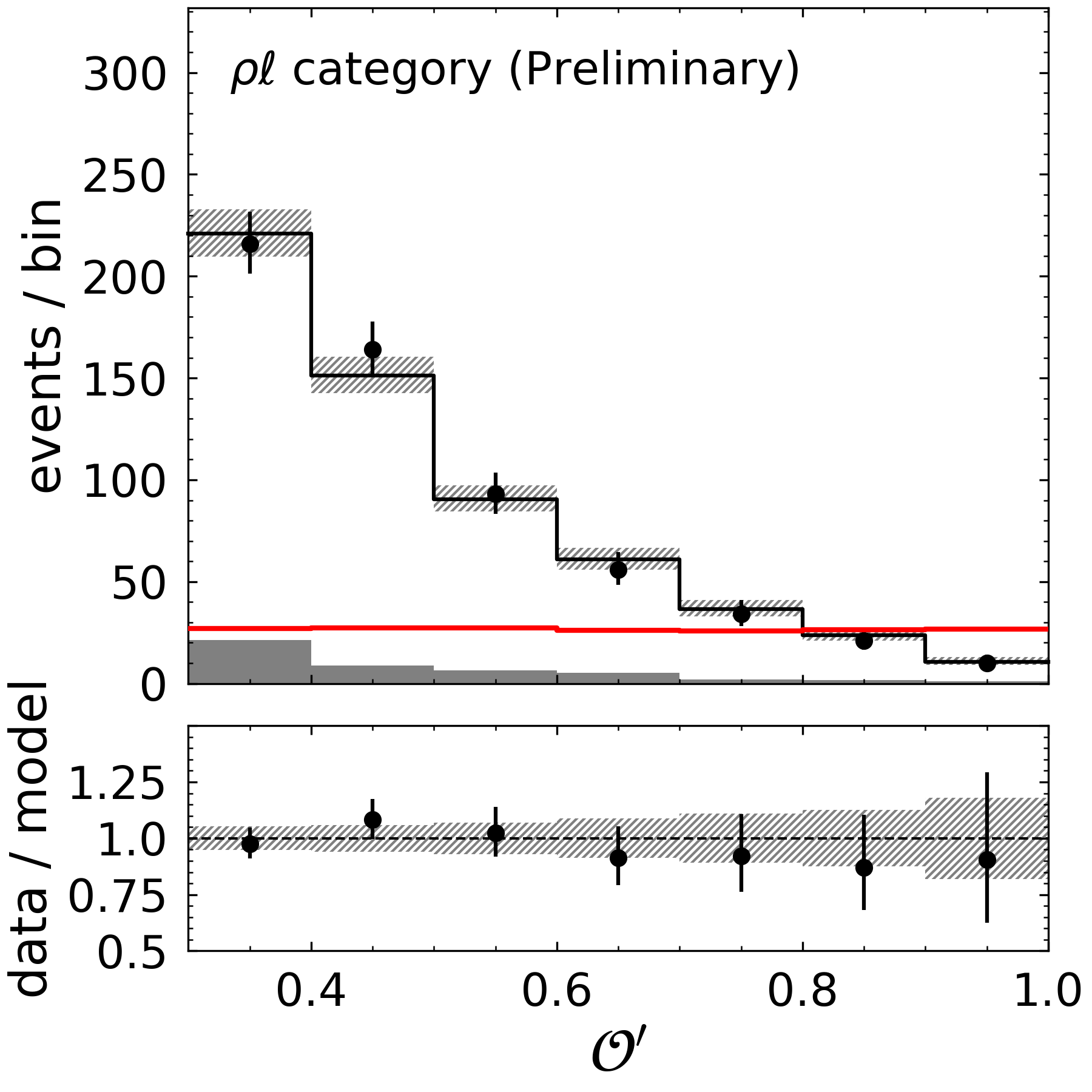}

    \end{subfigure}
    \begin{subfigure}{0.3\textwidth} 
        \includegraphics[width=\linewidth]{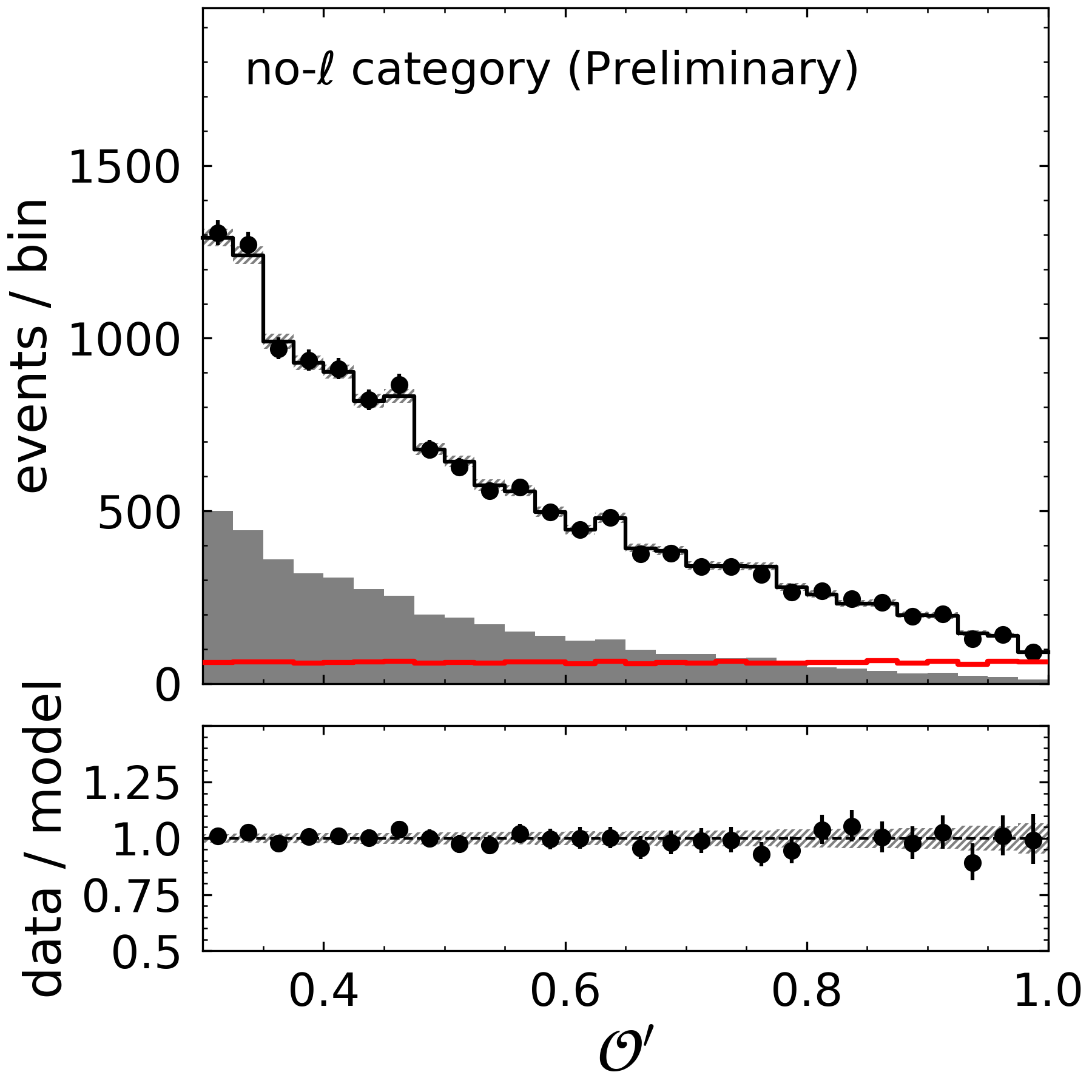}

    \end{subfigure}
    \begin{subfigure}{0.3\textwidth} 
        \includegraphics[width=\linewidth]{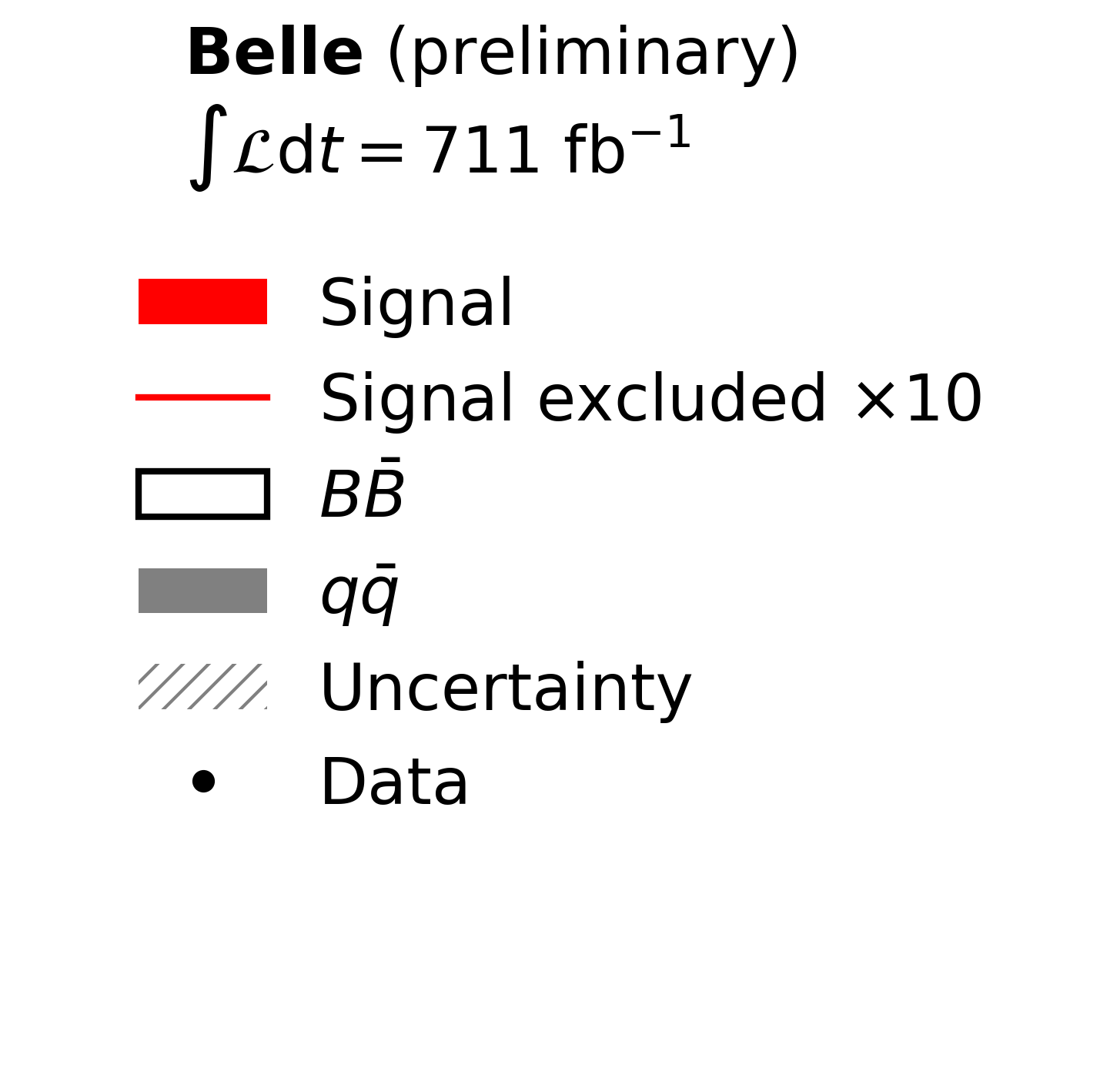}

    \end{subfigure}


    \begin{subfigure}{0.3\textwidth} 
        \includegraphics[width=\linewidth]{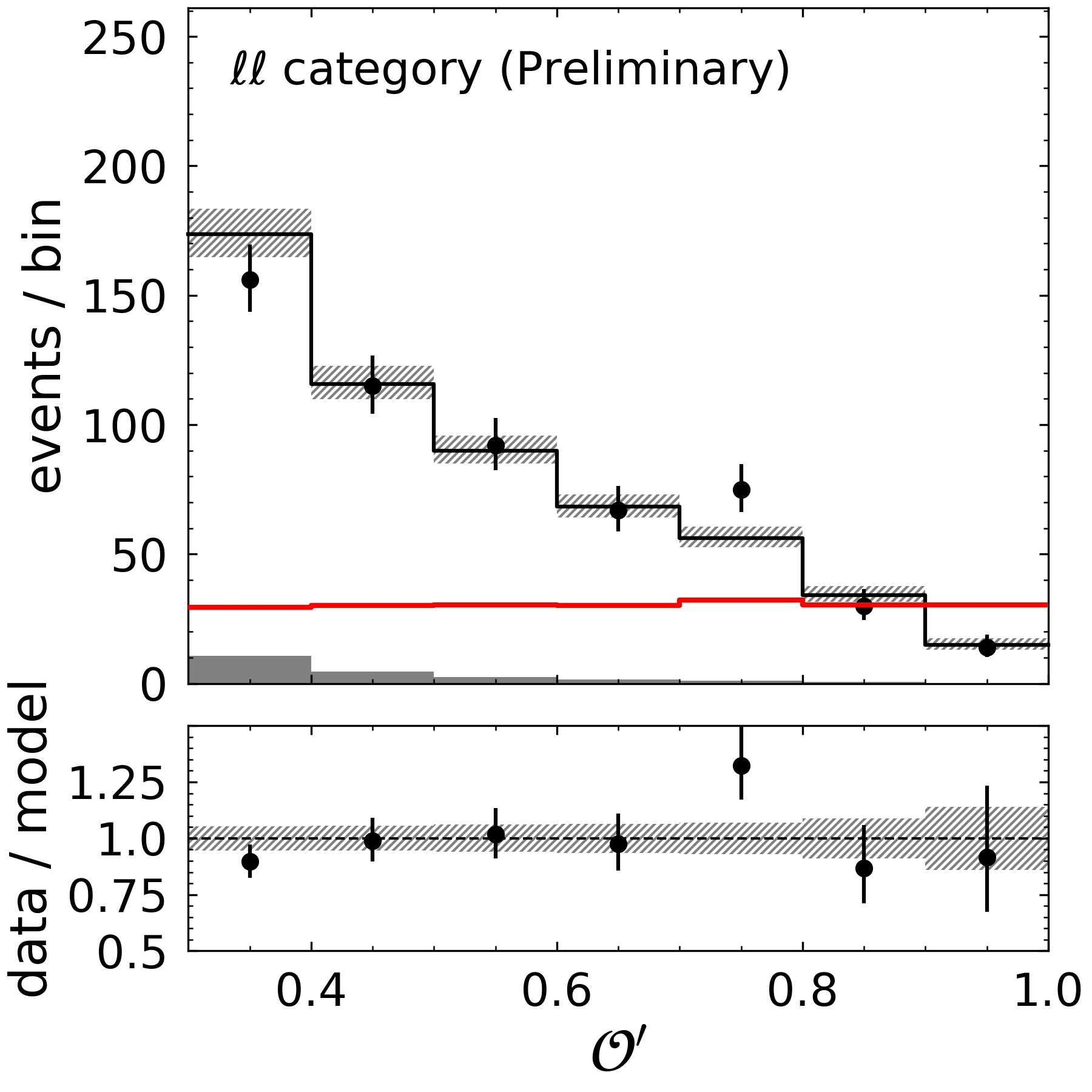}

    \end{subfigure}
    \begin{subfigure}{0.3\textwidth} 
        \includegraphics[width=\linewidth]{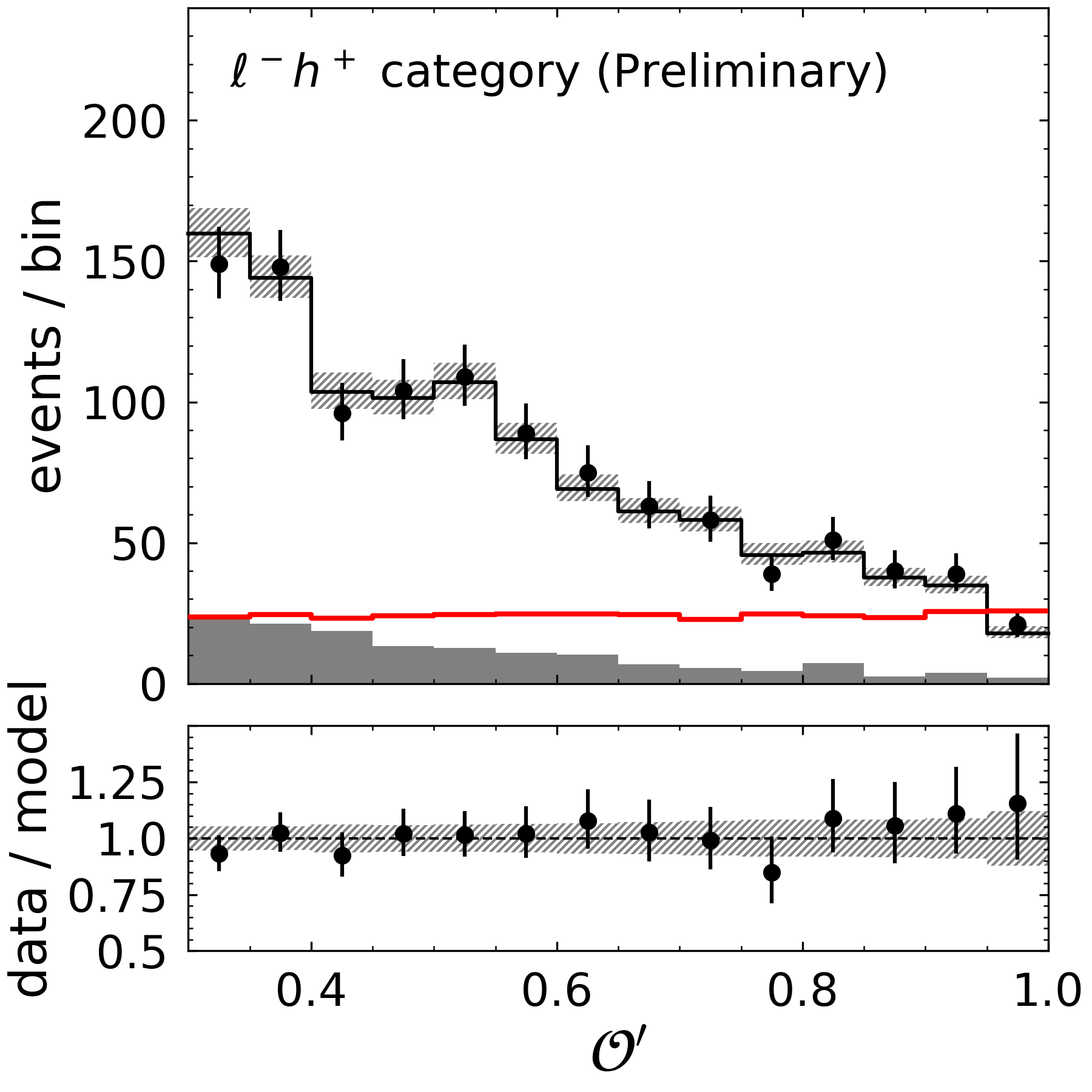}

    \end{subfigure}
    \begin{subfigure}{0.3\textwidth} 
        \includegraphics[width=\linewidth]{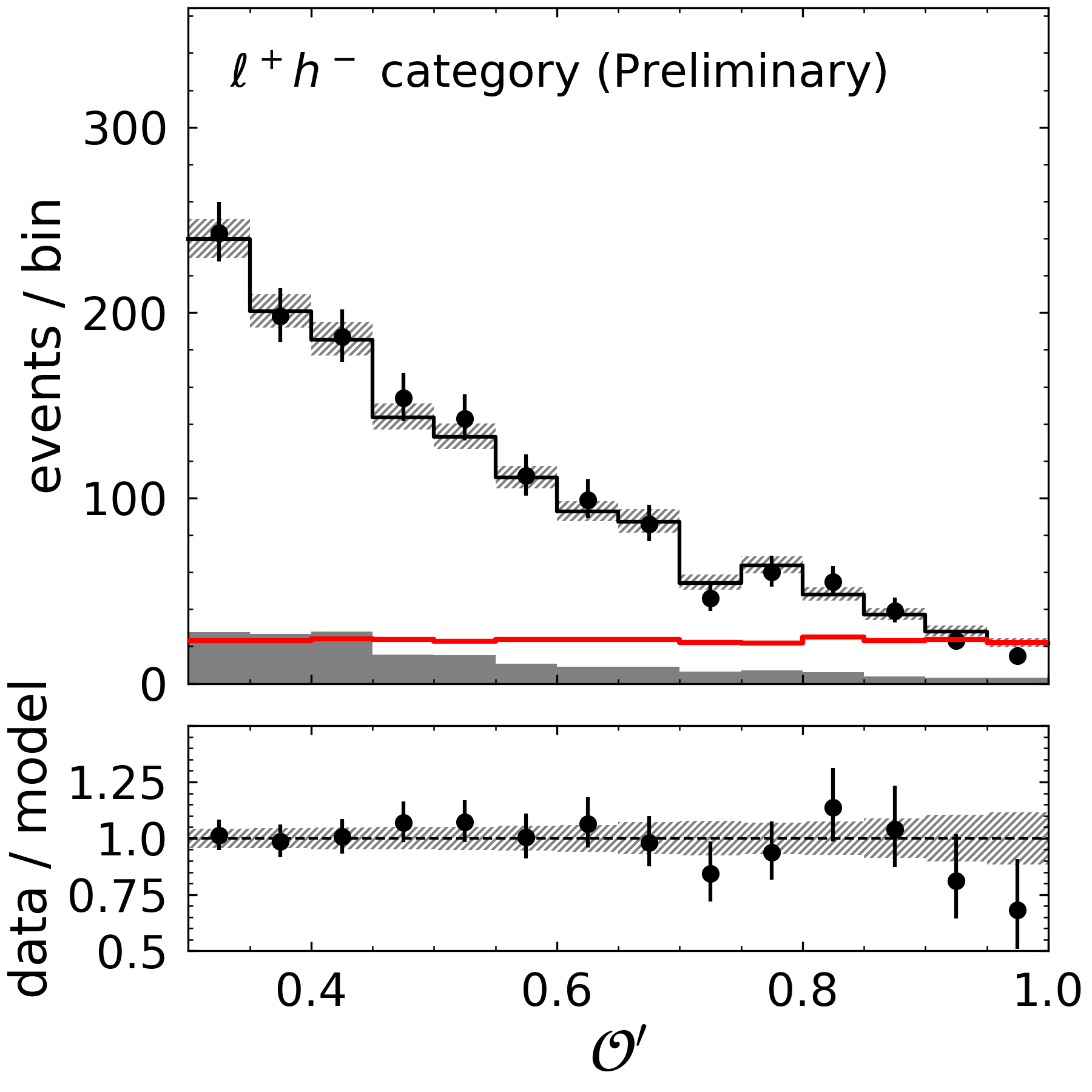}

    \end{subfigure}
    \begin{subfigure}{0.3\textwidth} 
        \includegraphics[width=\linewidth]{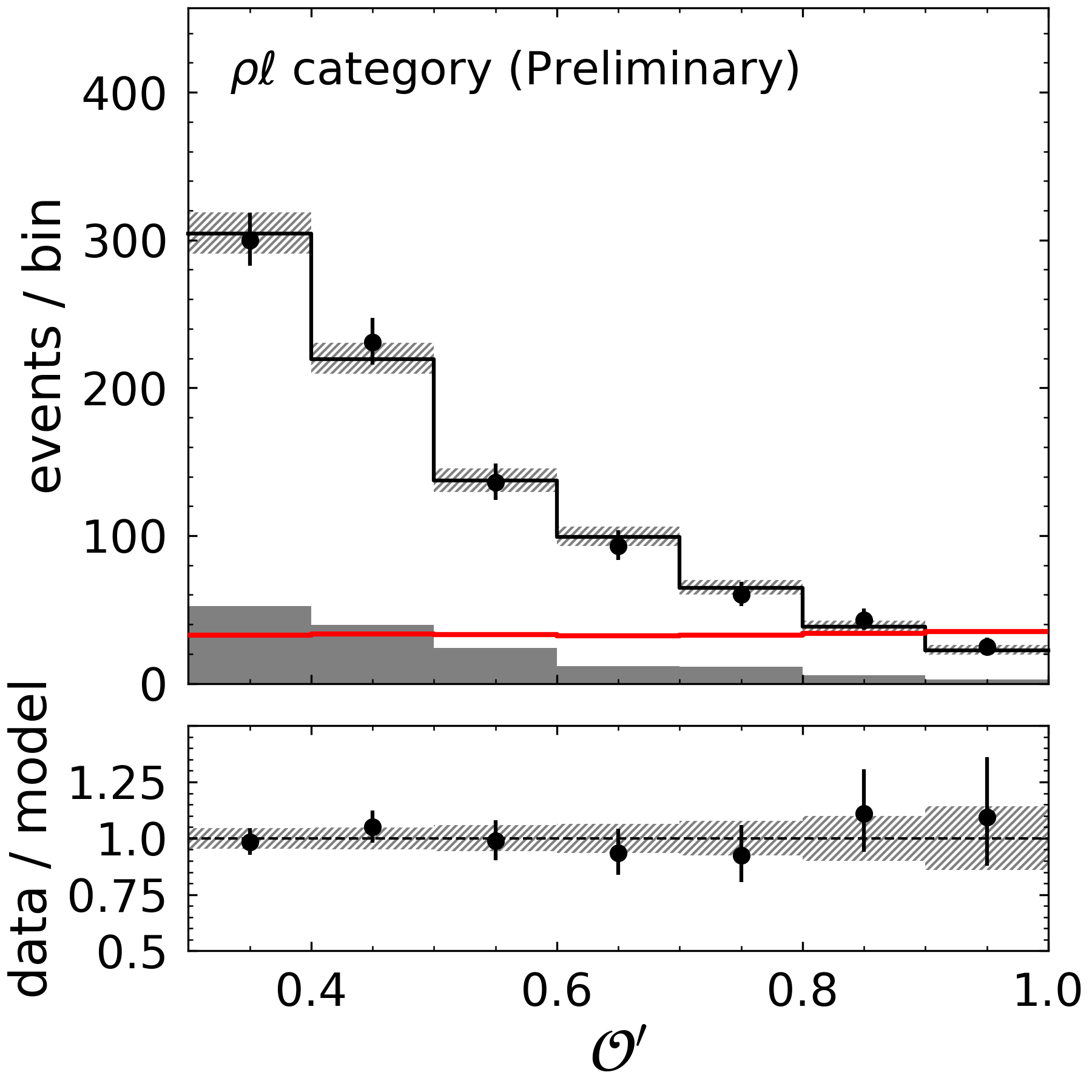}

    \end{subfigure}
    \begin{subfigure}{0.3\textwidth} 
        \includegraphics[width=\linewidth]{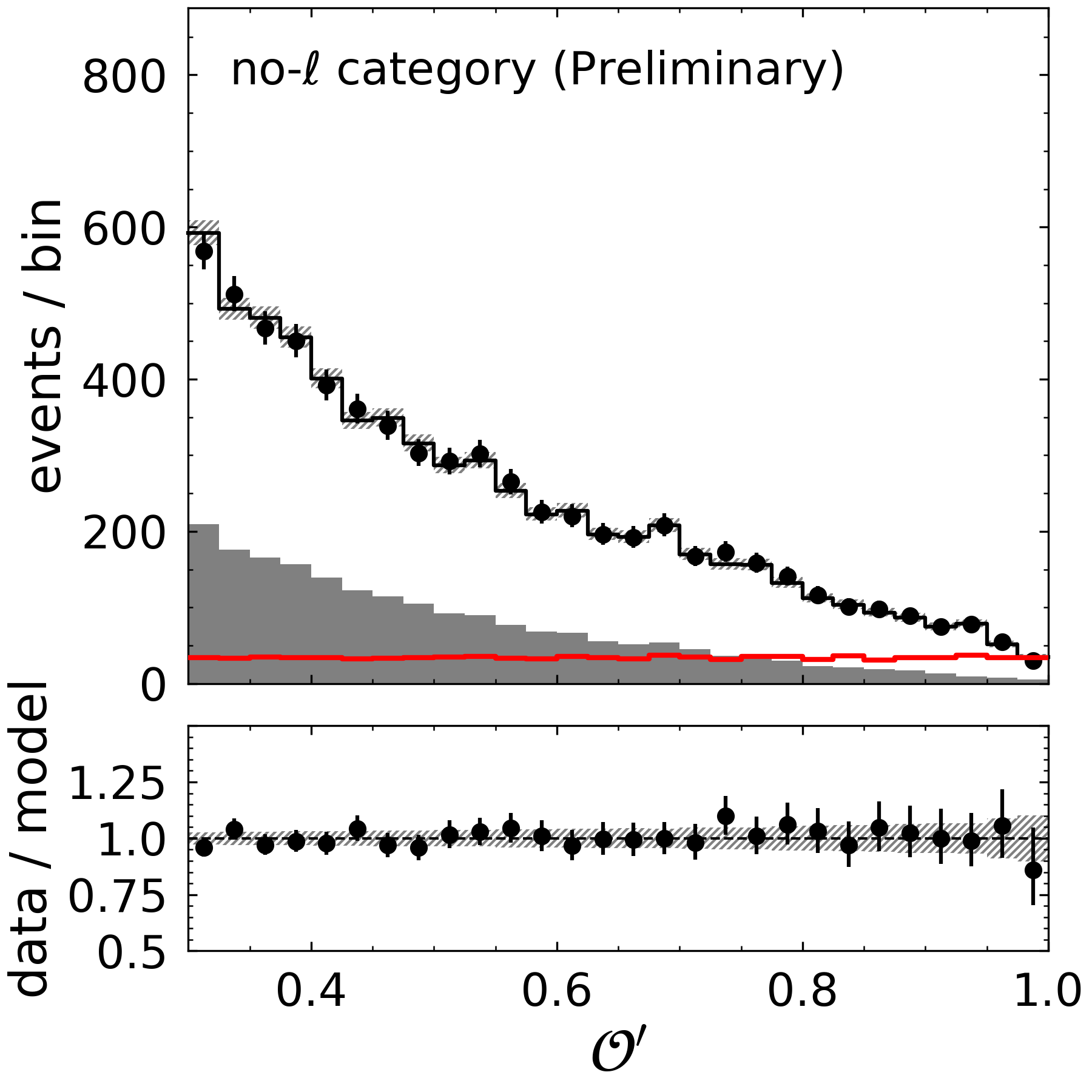}

    \end{subfigure}
    \begin{subfigure}{0.3\textwidth} 
        \includegraphics[width=\linewidth]{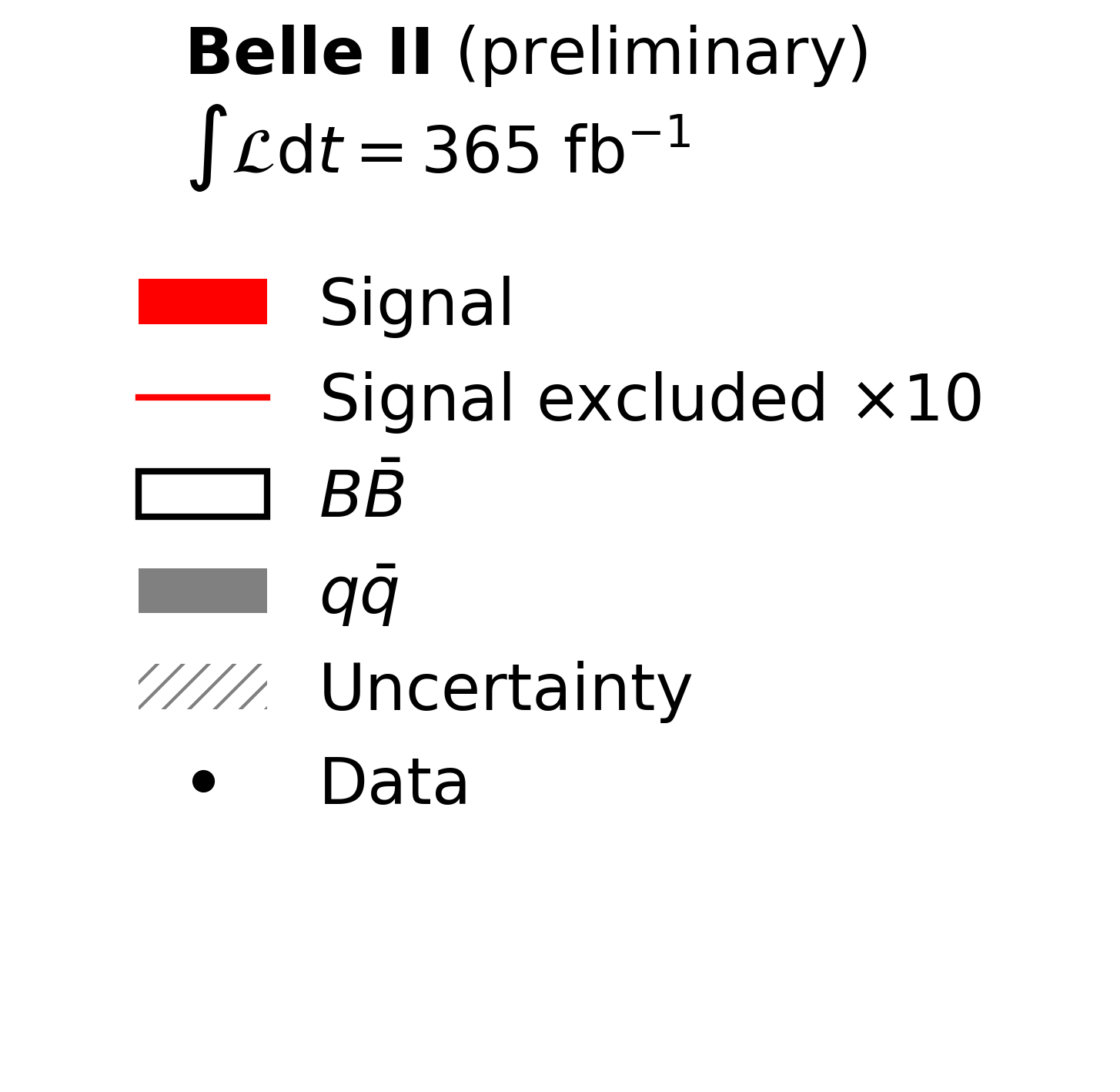}

    \end{subfigure}
    \caption{Distributions of $\mathcal{O}^\prime$ for candidates reconstructed in the Belle (upper five plots) and Belle~II (lower five plots) datasets together with the fit projections from a simultaneous fit. The red filled histogram (not visible) represents signal events at the branching fraction $\mathcal{B}(B^0 \to \KS \tau^+\tau^-) = {1.2} \times 10^{-4}$. The simulated background contributions from $e^+e^- \to q\bar{q}$ (grey filled histogram) and $e^+e^- \to \Upsilon(4S) \to B\bar{B}$ (open histogram) are stacked and the sum of components is compared with data.  The open red histogram (overlaid) illustrates the signal at a level ten times the exclusion limit on the branching fraction.  The panels below each distribution show the ratio between the data and model. The uncertainties on the histograms represent the total model uncertainty as determined by the fit.\label{fig:fit}}
\end{figure}

The signal extraction fit is performed simultaneously for the five signal categories and both the Belle and Belle~II datasets, using the binned distributions of the transformed BDT output observable $\mathcal{O}^\prime$ in the range $[0.3, 1.0]$. Both statistical uncertainties and systematic effects are incorporated directly into the likelihood model used for the fit.
The precision of the measurement is limited by the statistical uncertainty. 
While the total systematic uncertainty is similar in size, the second dominant source arises from the finite size of the simulated sample and is therefore statistical in nature.
The largest systematic contribution stems from the limited knowledge of the signal-side $B$ composition. Fit projections\footnote{Numerical data are available as HEPData~\cite{hepdata}.} in $\mathcal{O}^\prime$ show good agreement with the observed data as shown in Fig.~\ref{fig:fit}. 

\begin{figure}[!ht]
    \centering
    \begin{subfigure}{0.3\textwidth} 
        \includegraphics[width=\linewidth]{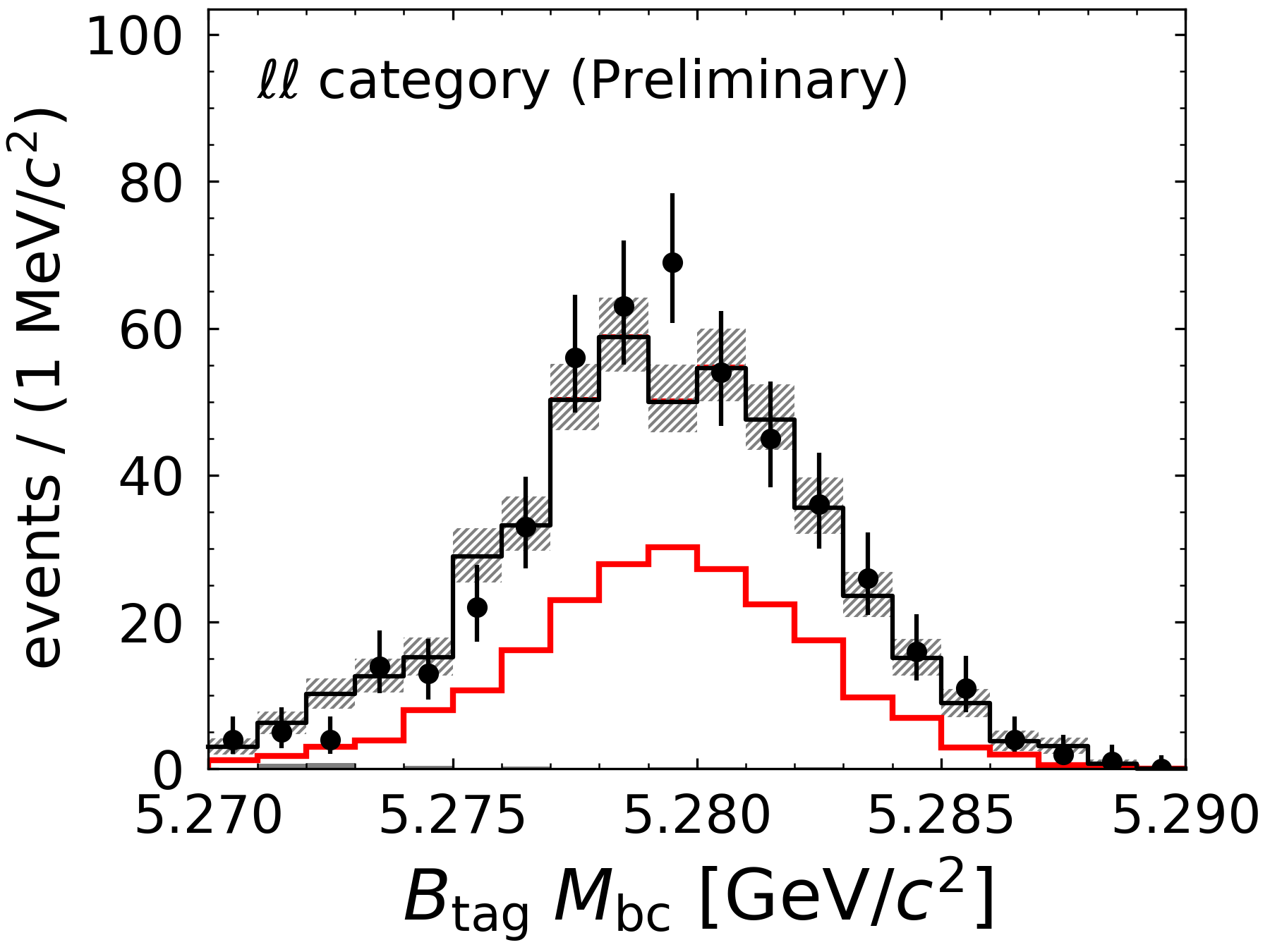}

    \end{subfigure}
    \begin{subfigure}{0.3\textwidth} 
        \includegraphics[width=\linewidth]{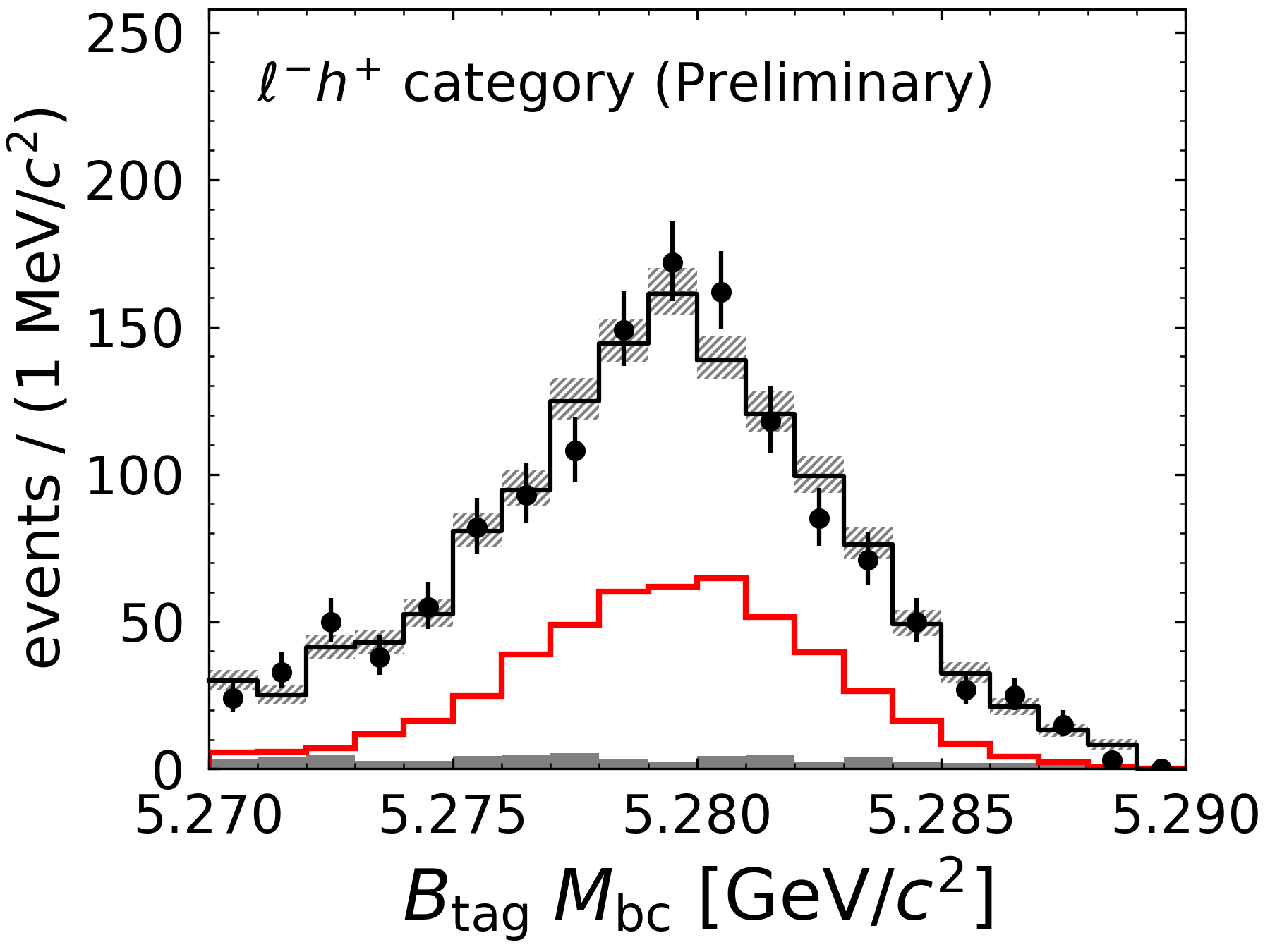}

    \end{subfigure}
    \begin{subfigure}{0.3\textwidth} 
        \includegraphics[width=\linewidth]{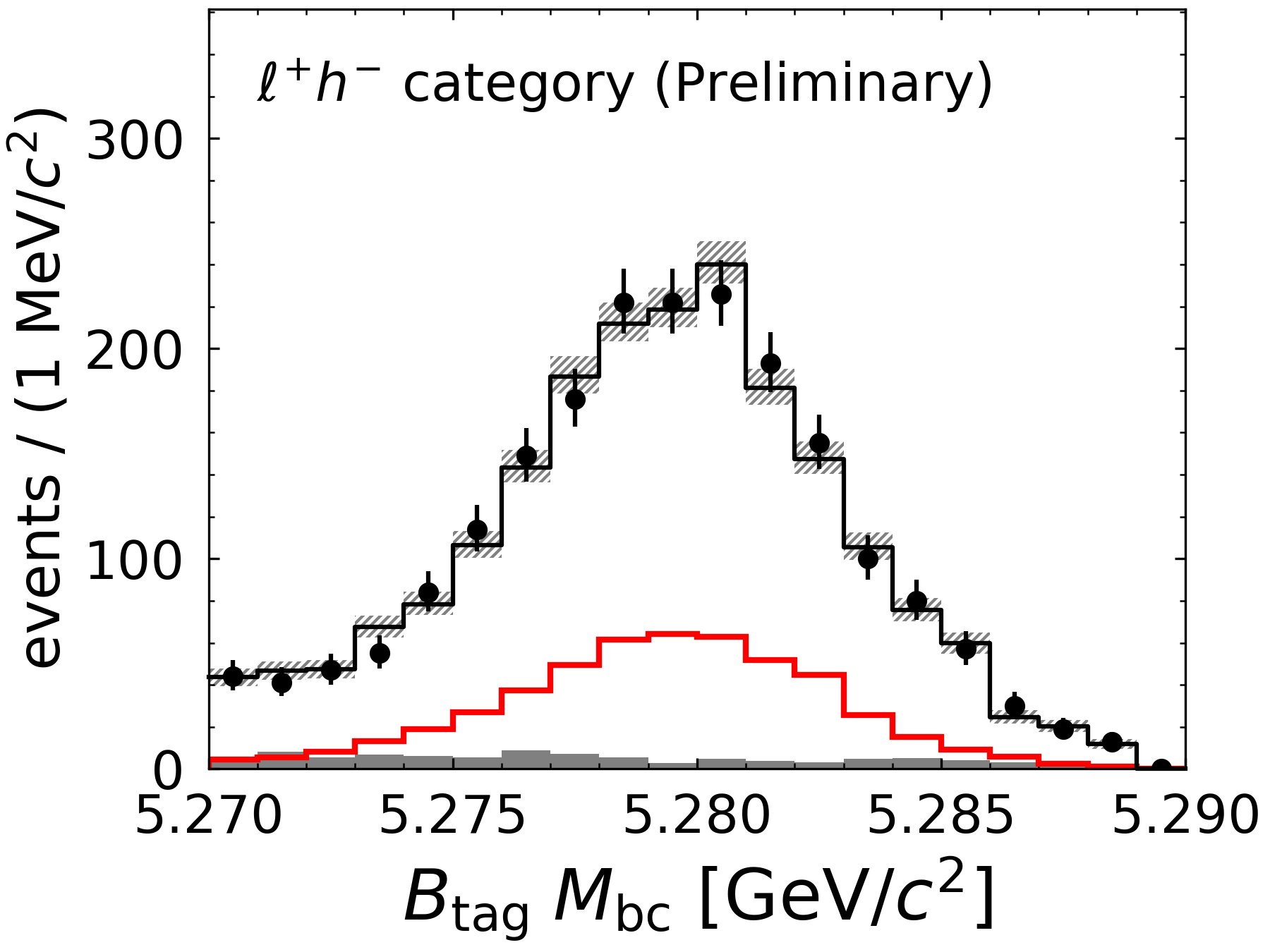}

    \end{subfigure}
    \begin{subfigure}{0.3\textwidth} 
        \includegraphics[width=\linewidth]{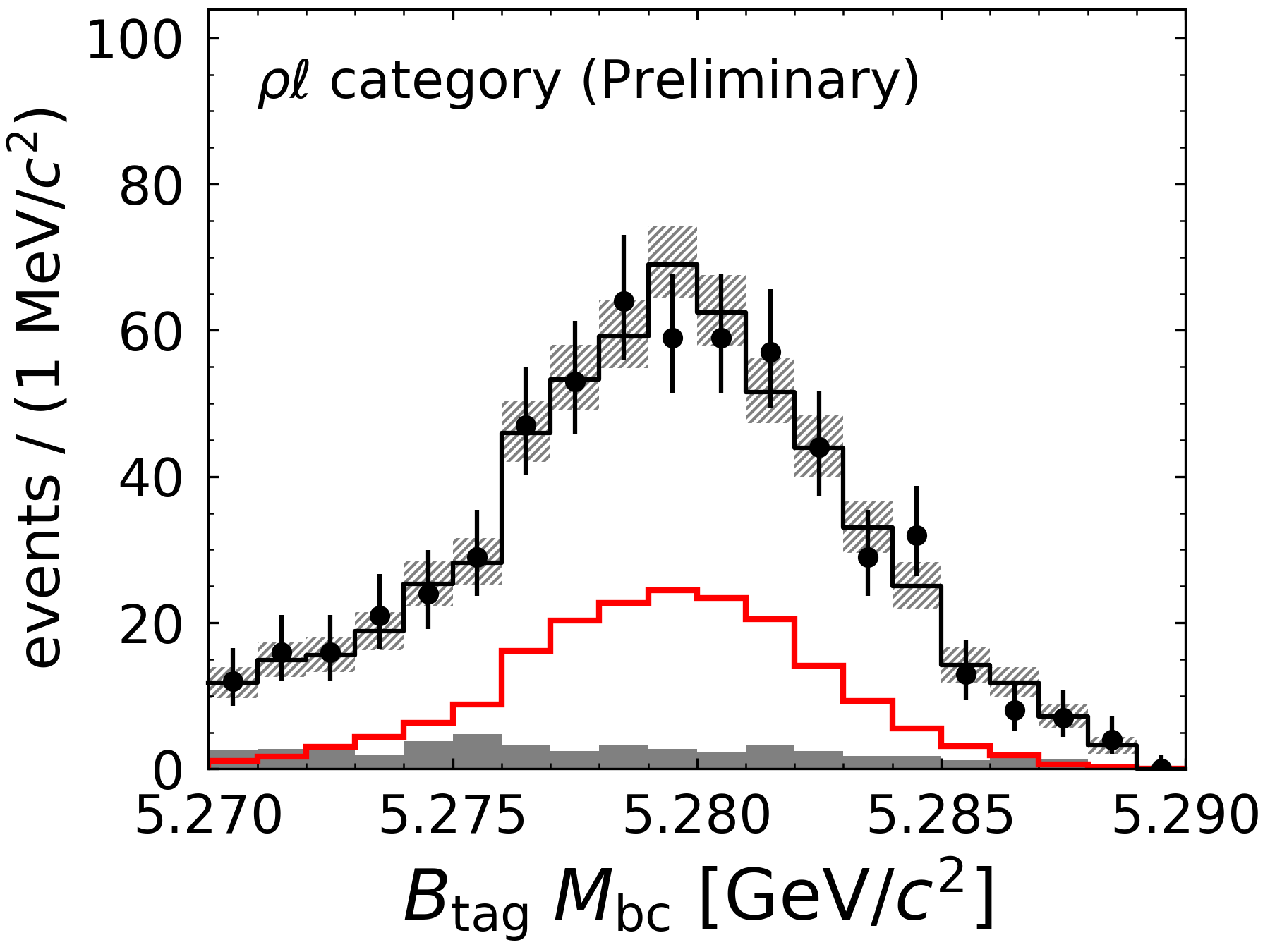}

    \end{subfigure}
    \begin{subfigure}{0.3\textwidth} 
        \includegraphics[width=\linewidth]{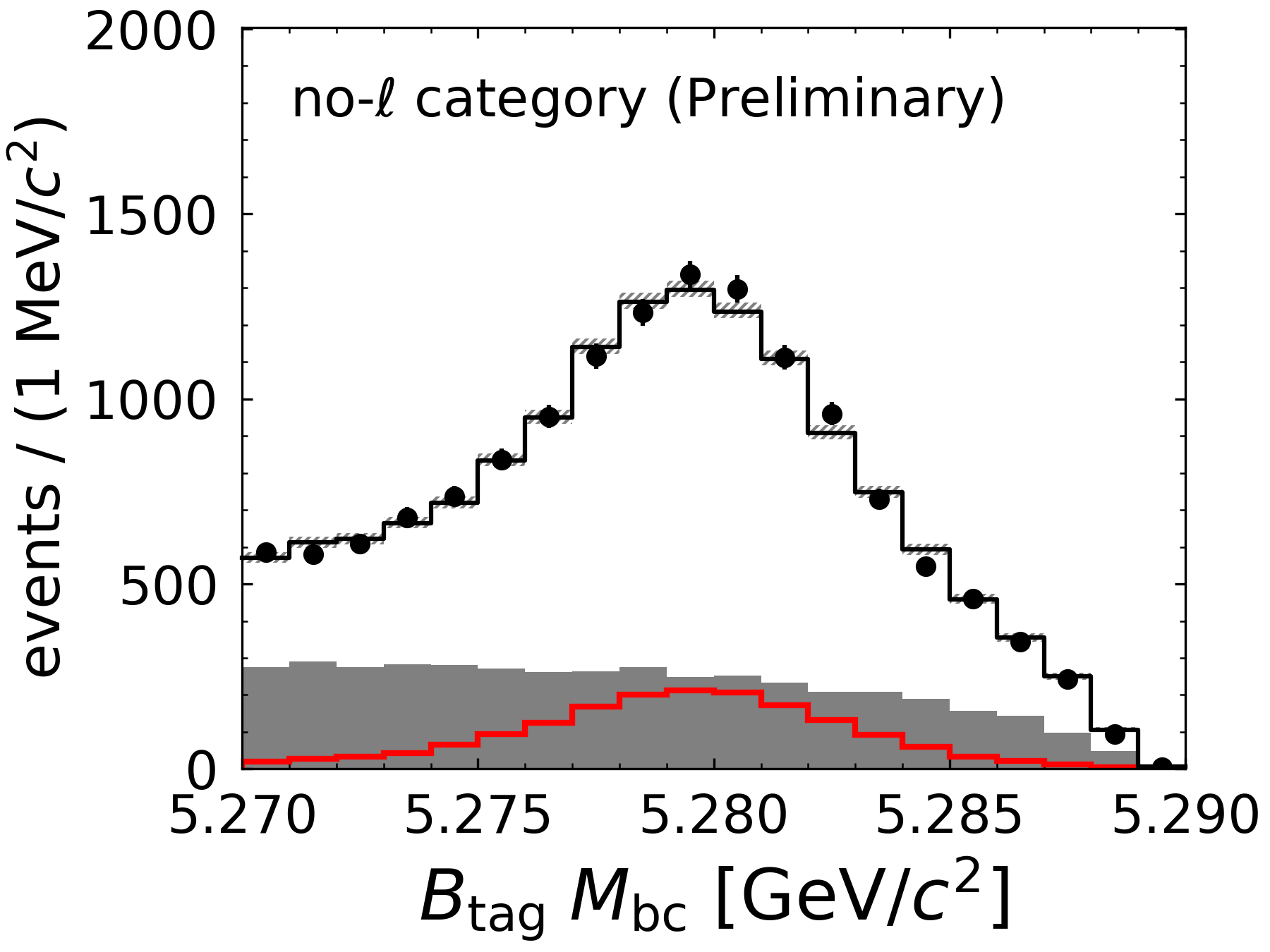}

    \end{subfigure}
    \begin{subfigure}{0.3\textwidth} 
        \includegraphics[width=\linewidth]{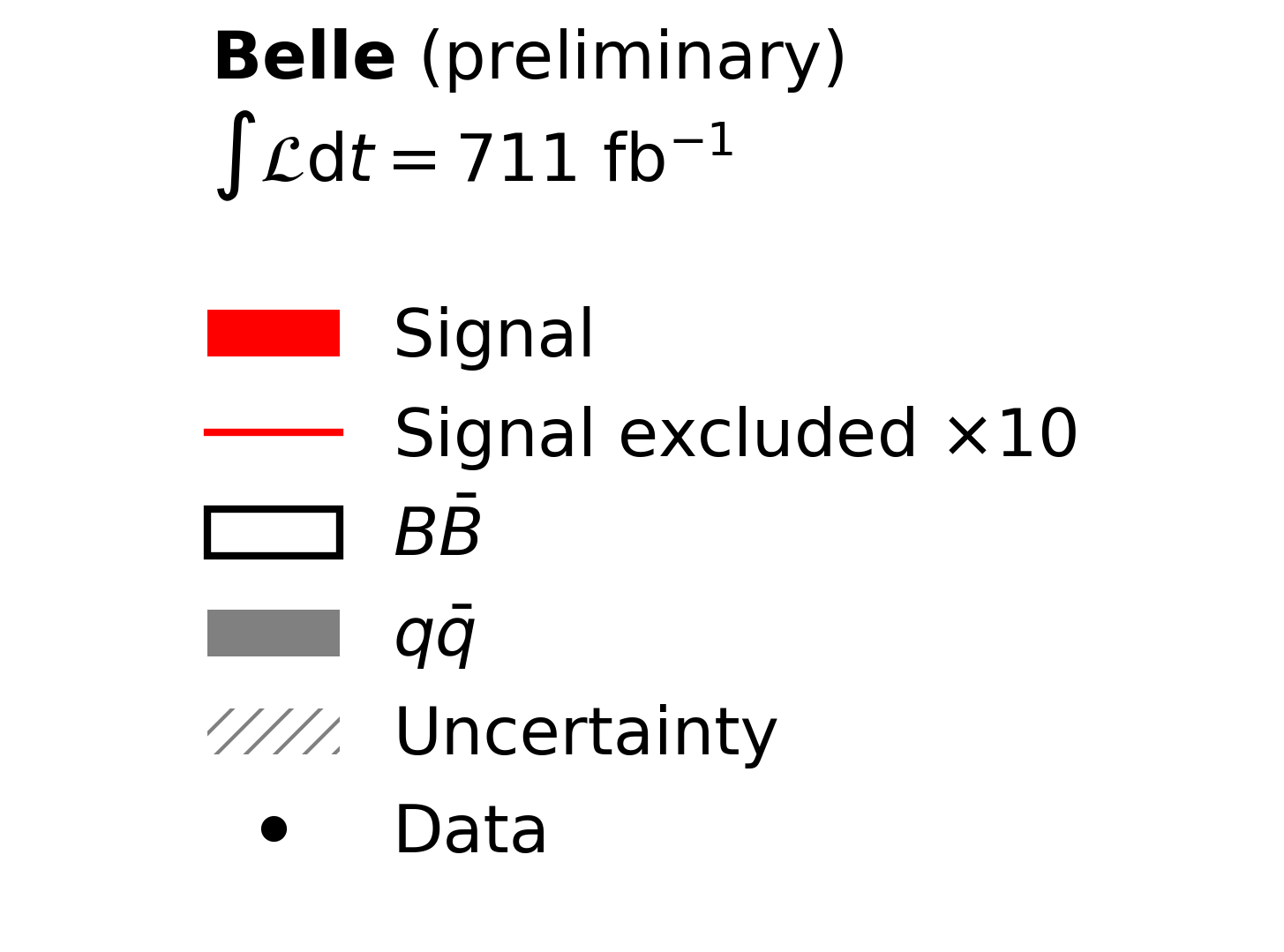}

    \end{subfigure}


    \begin{subfigure}{0.3\textwidth} 
        \includegraphics[width=\linewidth]{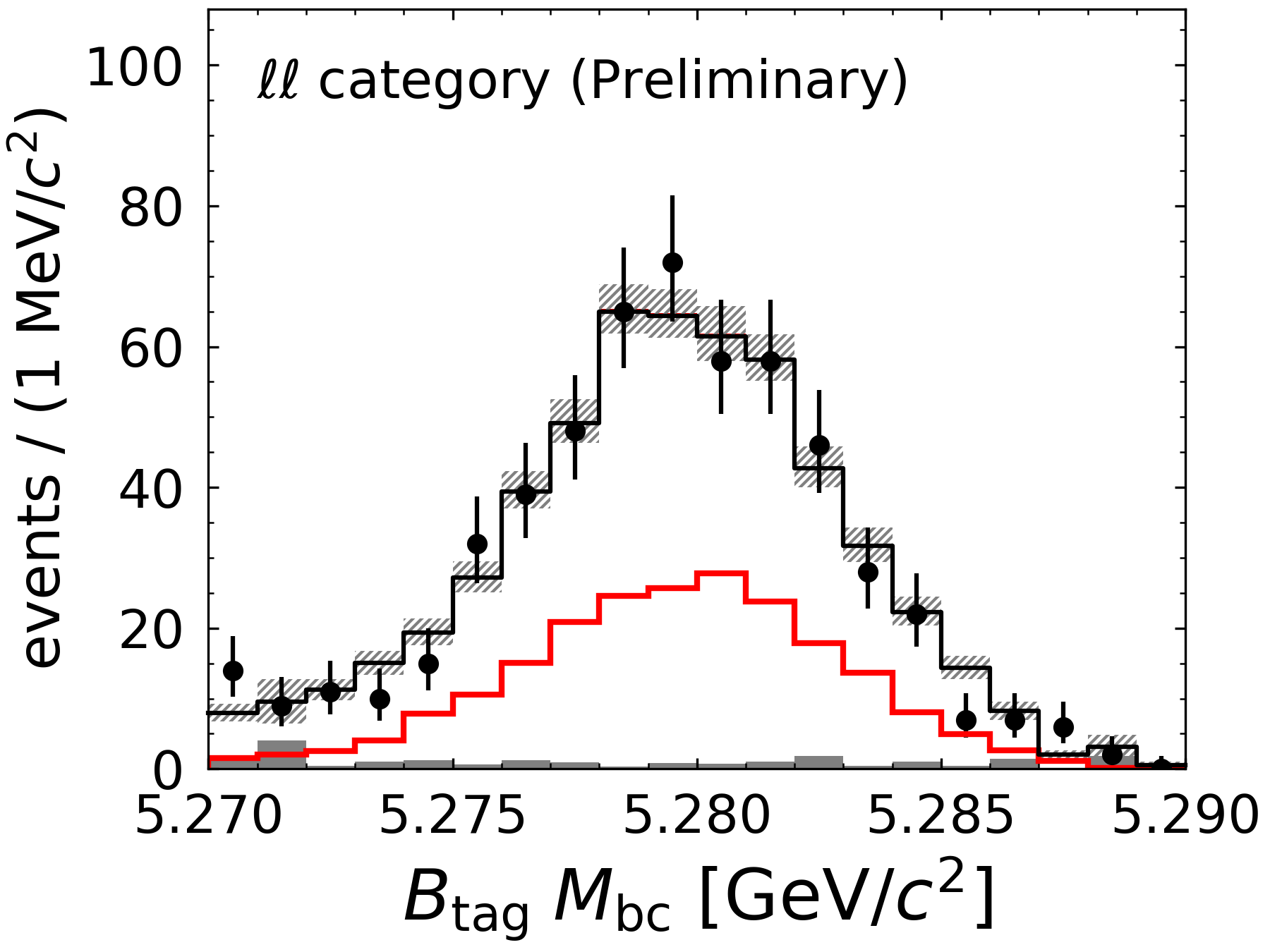}

    \end{subfigure}
    \begin{subfigure}{0.3\textwidth} 
        \includegraphics[width=\linewidth]{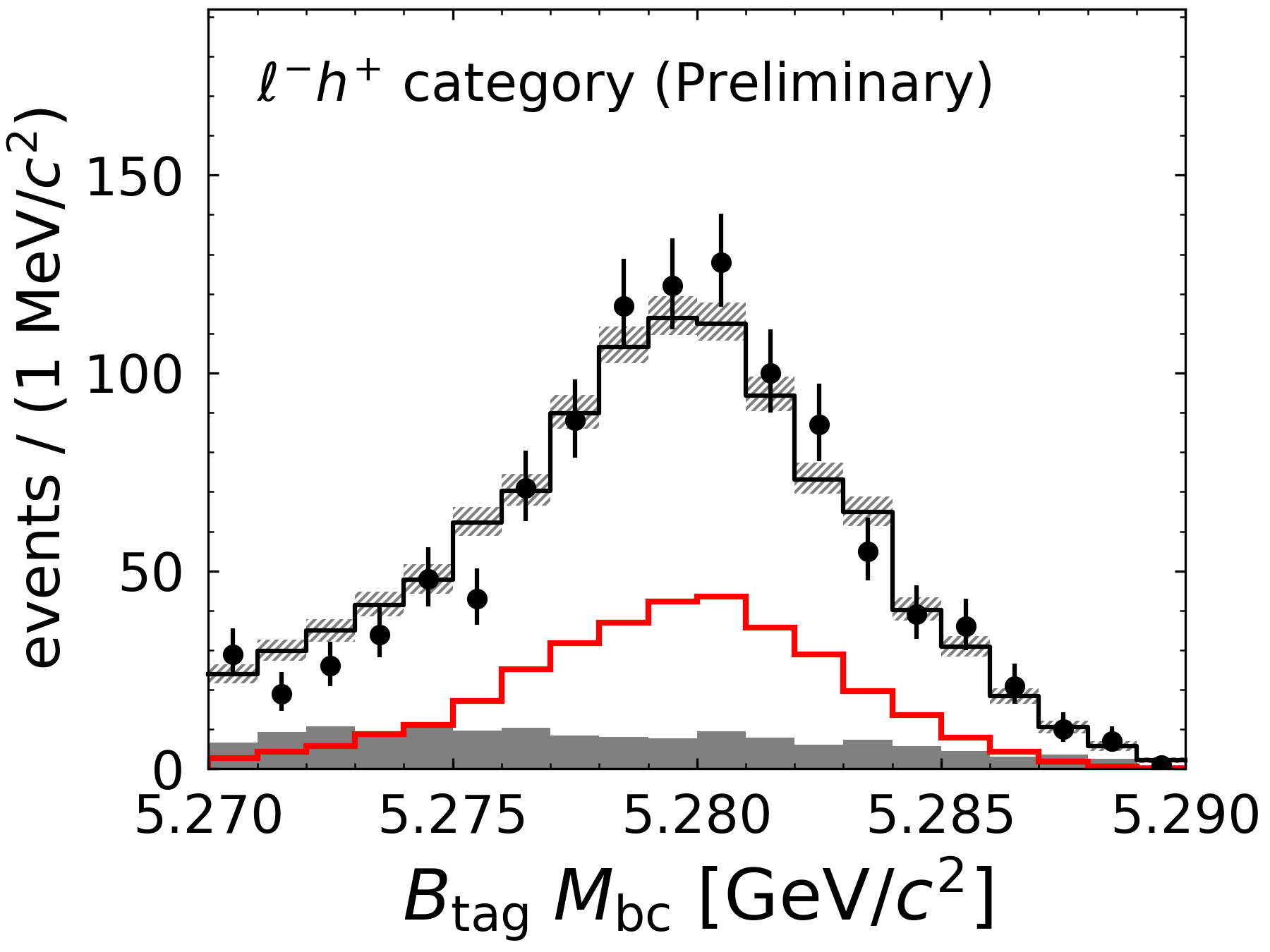}

    \end{subfigure}
    \begin{subfigure}{0.3\textwidth} 
        \includegraphics[width=\linewidth]{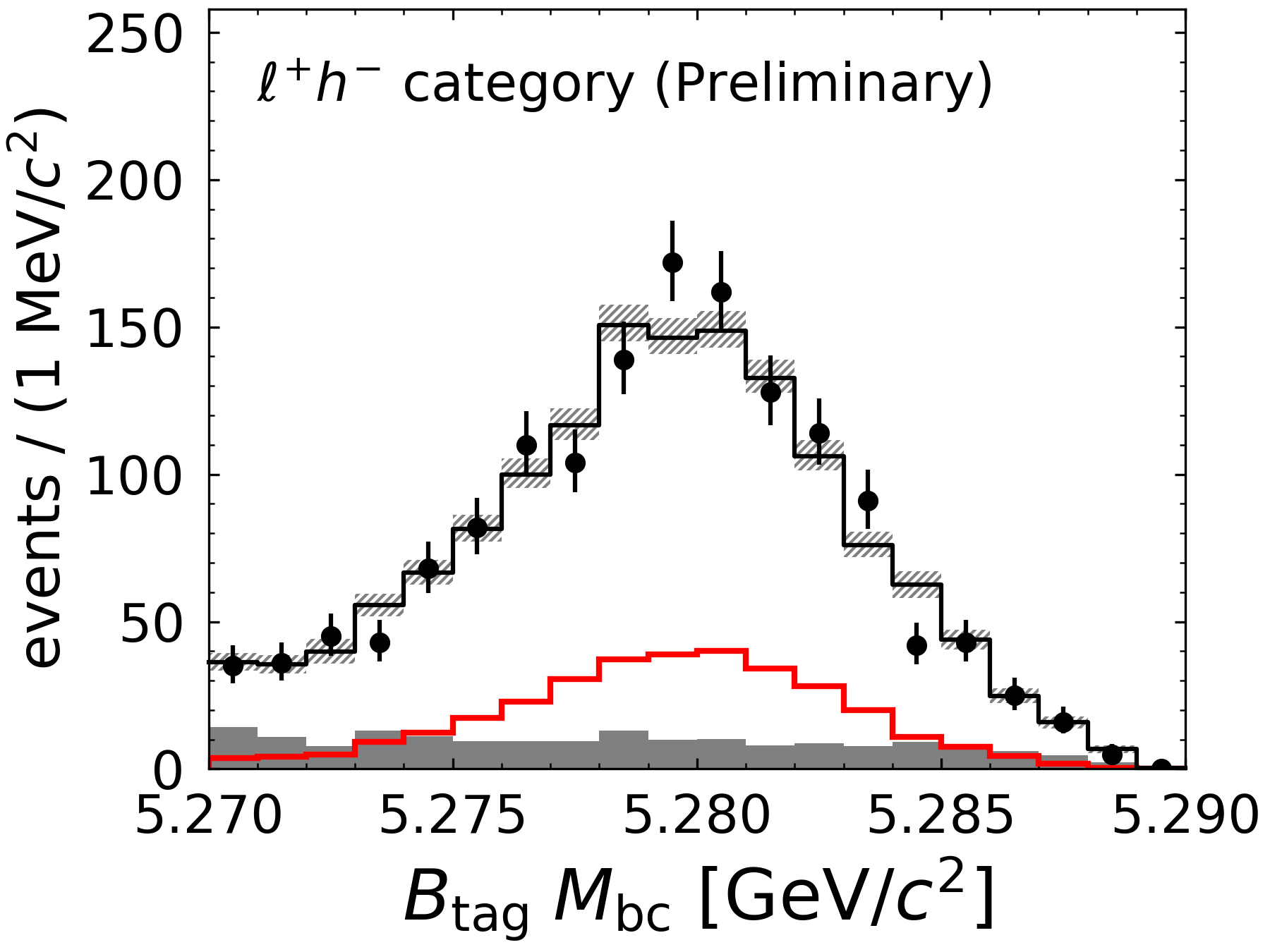}

    \end{subfigure}
    \begin{subfigure}{0.3\textwidth} 
        \includegraphics[width=\linewidth]{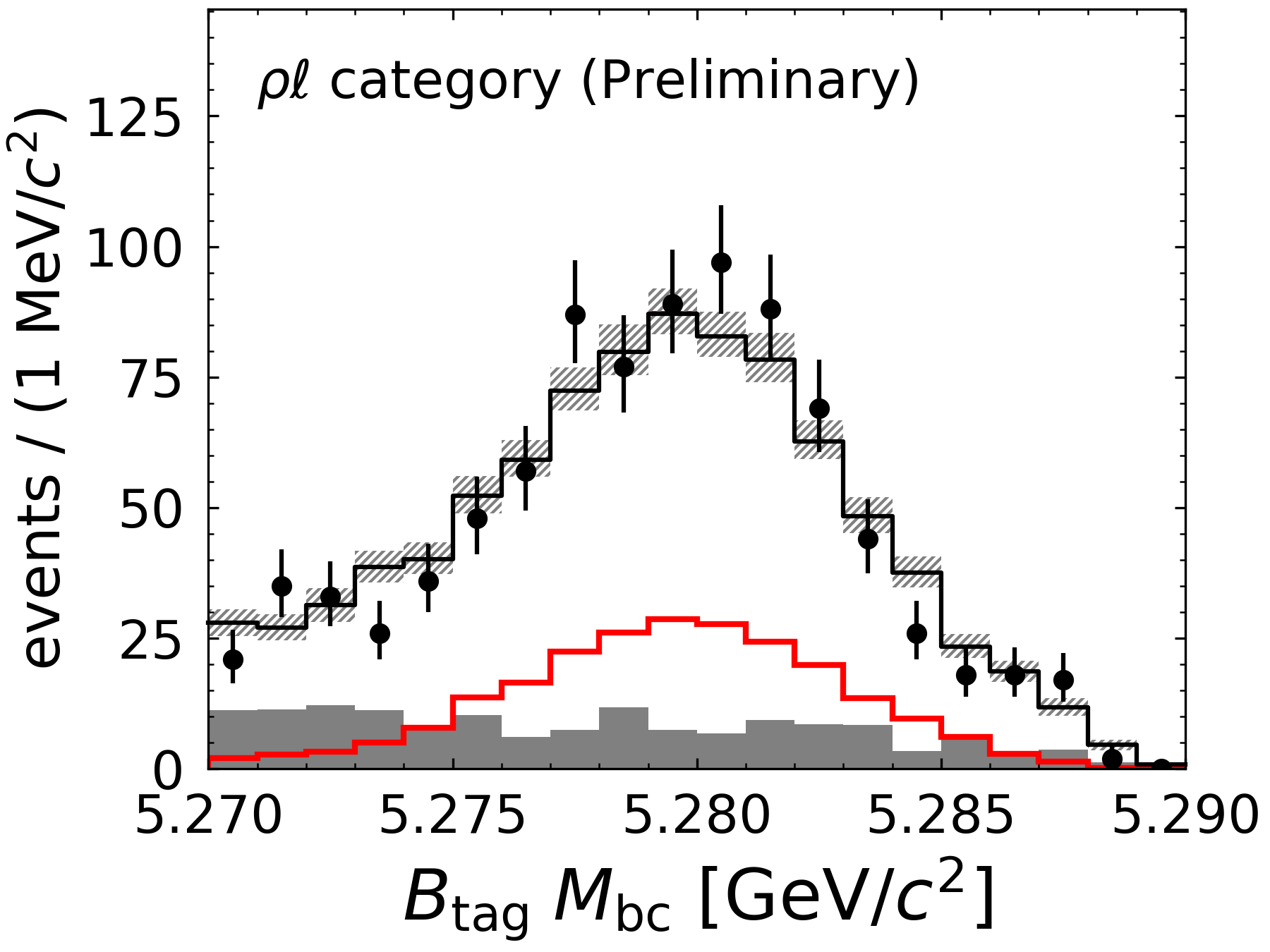}

    \end{subfigure}
    \begin{subfigure}{0.3\textwidth} 
        \includegraphics[width=\linewidth]{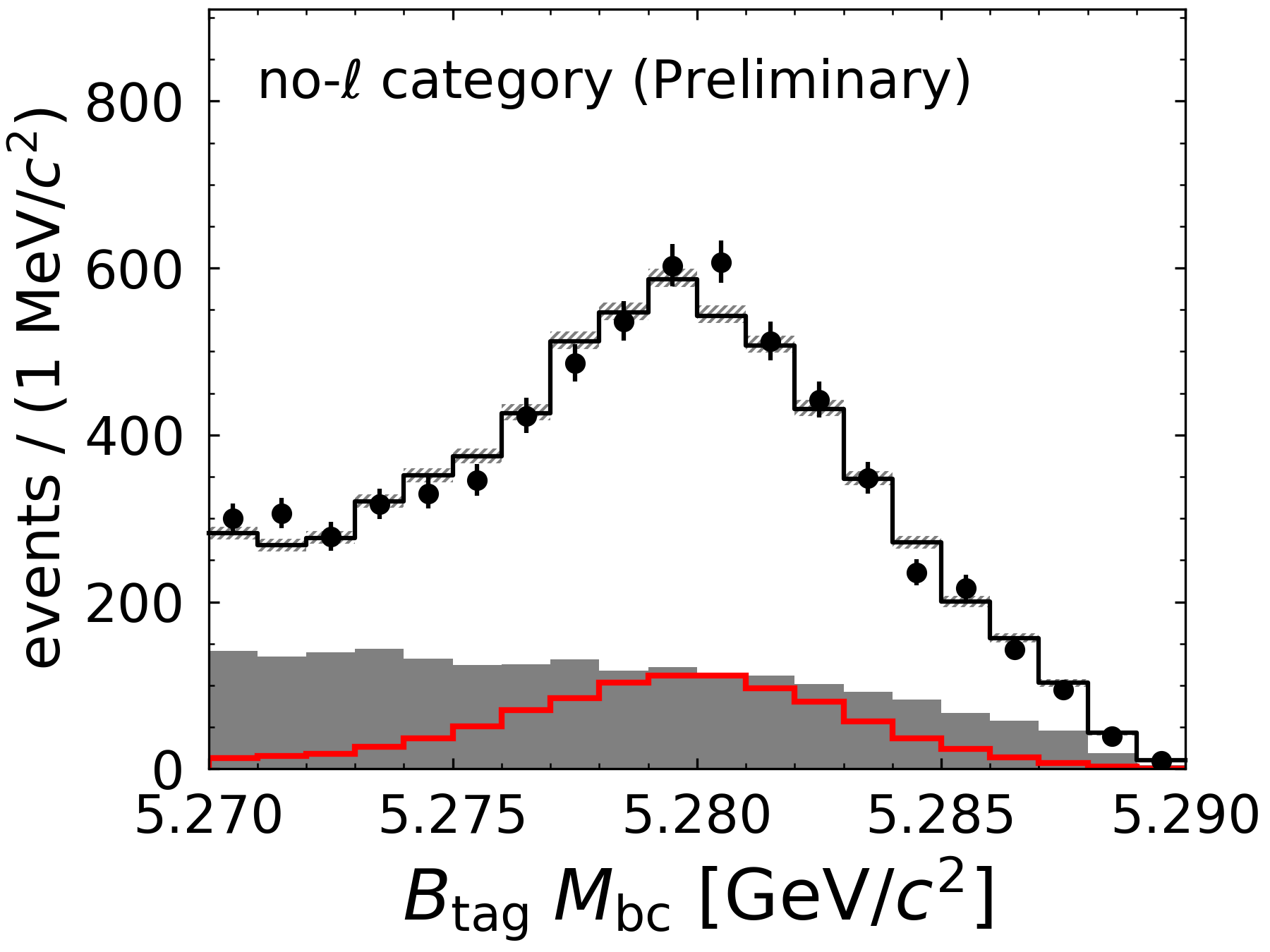}

    \end{subfigure}
    \begin{subfigure}{0.3\textwidth} 
        \includegraphics[width=\linewidth]{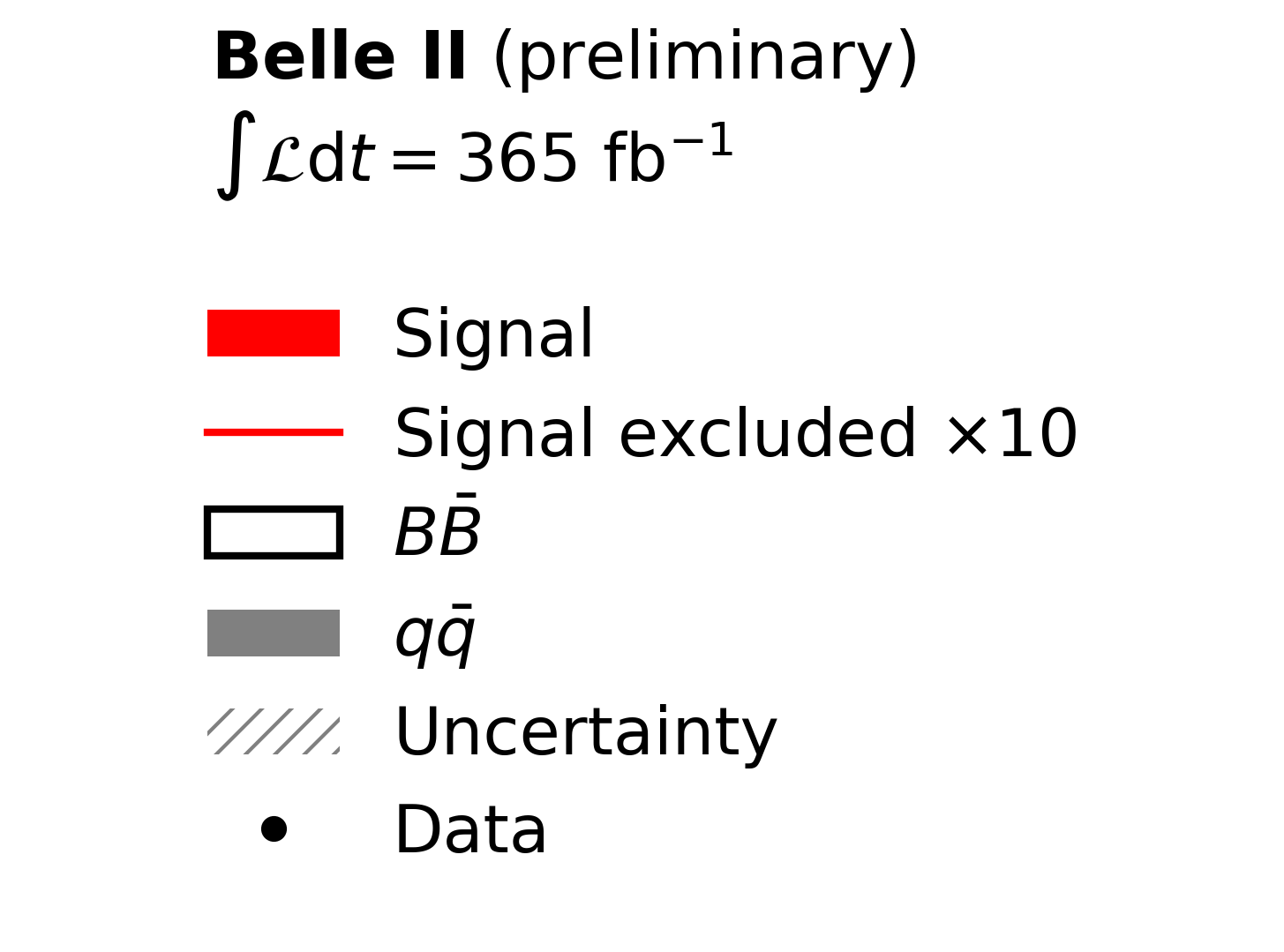}
    \end{subfigure}
    \caption{Distributions of $M_{\rm bc}$ for tag-side $B$ candidates reconstructed in the Belle (upper five plots) and Belle~II (lower five plots) datasets. Histograms from simulated samples are shown with yields rescaled to account for the simultaneous fit to the $\mathcal{O}^\prime$ observable. The red filled histogram (barely visible) represents the observed signal events at the  branching fraction $\mathcal{B}(B^0 \to \KS \tau^+\tau^-) = {1.2} \times 10^{-4}$.  The simulated background contributions from $e^+e^- \to \Upsilon(4S) \to B\bar{B}$ (open histogram) and  $e^+e^- \to q\bar{q}$ (grey filled histogram) are stacked and the sum of components is compared with data.  The open red histogram (overlaid) illustrates the signal at a level ten times the exclusion limit on the branching fraction. The uncertainties shown on the histograms reflect the statistical uncertainties of the simulation. \label{fig:mbc_mbc}}
\end{figure}

The fit observable $\mathcal{O}^\prime$ is constructed from observables related to the signal-side $B$ candidate, the full event, and its unreconstructed part. Therefore, the information from the tag-side $B$ candidate is largely uncorrelated with $\mathcal{O}^\prime$ and is suitable for post-fit validation. As an example, Fig.~\ref{fig:mbc_mbc} shows the beam-constrained mass distributions of the tag-side $B$, with the templates scaled according to the fit results in each $\mathcal{O}^\prime$ bin. The model provides a good description of the data, reproducing the relative composition of peaking and combinatorial tag candidates across all the categories.

We measure the branching fraction in the entire allowed $q^2$ region,
\begin{equation*}
\mathcal{B}(B^0 \to \KS \tau^+\tau^-) = ({1.2} \pm {3.3}  \pm {2.9} ) \times 10^{-4},
\end{equation*}
 where the first uncertainty is statistical and the second is systematic. The results of separate fits to either Belle or Belle~II data are $\mathcal{B}(B^0 \to \KS \tau^+\tau^-) = ({0.7} \pm 4.2\pm {3.6}) \times 10^{-4}$ and $\mathcal{B}(B^0 \to \KS \tau^+\tau^-) = (2.4 \pm 5.0 \pm 4.4)\times10^{-4}$, respectively. 
 
  The observed [expected] exclusion limit is evaluated at the 90\% confidence level yielding
\begin{equation*}
\mathcal{B}(B^0 \to \KS \tau^+\tau^-) < {8.3} \,{[7.4]} \times 10^{-4}\,.
\end{equation*}

This represents the only limit on this decay to date, offering sensitivity to BSM scenarios that predict enhancements in $b \to s \tau^+ \tau^-$ transitions involving pseudoscalar kaons. 



We combine this result with the recent measurement of $B^+ \to K^+ \tau^+\tau^-$~\cite{butokplustautauB1B2} under the assumption of isospin symmetry. The combination is performed by constructing a joint likelihood from the binned signal-extraction models of the two measurements, with a single shared parameter of interest corresponding to the isospin-averaged branching fraction $\mathcal{B}(B\to K\tau^+\tau^-)$. This parameter is related to the charged and neutral decay modes as 
$
\mathcal{B}(B\to K\tau^+\tau^-) \equiv \mathcal{B}(B^+ \to K^+ \tau^+\tau^-) =2\,(\tau_{B^+}/\tau_{B^0})\, \mathcal{B}(B^0 \to \KS \tau^+\tau^-),
$
where $\tau_{B^+}/\tau_{B^0}=1.076\pm0.004$~\cite{ParticleDataGroup:2024cfk} and $\tau_B$ represents the $B$-meson lifetime.
The factor of two accounts for the relation $\mathcal{B}(B^0\to K^0\tau^+\tau^-)= 2\,\mathcal{B}(B^0\to \KS\tau^+\tau^-)$. The uncertainty on the lifetime ratio is included as an additional normalisation uncertainty, and correlations between systematic uncertainties common to the two analyses are taken into account. The fit and limit extraction are then performed using the combined likelihood.

As a result, we measure 
$\mathcal{B}(B\to K \tau^+ \tau^-) = (0.6\pm2.8)\times 10^{-4}$. The inclusion of the $B^0\to\KS\tau^+\tau^-$ result improves the observed [expected] upper limit by about 5\% [10\%] relative to the limit achieved with $B^+ \to K^+ \tau^+\tau^-$ alone. The resulting observed [expected] upper limit is $\mathcal{B}(B\to K \tau^+ \tau^-)<5.4\,[5.0]\times 10^{-4}$ at the 90\% confidence level.

\section{Summary} \label{sec:summary}
We perform the first search for the rare decay $B^0 \to \KS \tau^+\tau^-$ and use the Belle and Belle~II datasets of $772 \times 10^6$ and $387 \times 10^6$ $\Upsilon(4S)$ events, collected at the KEKB and SuperKEKB colliders, respectively. 
The analysis tags one $B$ meson in a hadronic decay mode, reconstructs the $\tau$ leptons through decay modes with a single charged particle, and uses multivariate classifiers tailored for each of five mutually exclusive $\tau^+\tau^-$ final state categories. Signal extraction is performed through a simultaneous binned maximum likelihood fit to the classifier outputs in all reconstructed $\tau^+\tau^-$ categories in the Belle and Belle~II datasets.

We do not observe a significant $B^0 \to \KS \tau^+\tau^-$ signal and set an upper limit $\mathcal{B}(B^0 \to \KS \tau^+\tau^-)<{8.3} \times 10^{-4}$ at the 90\% confidence level. We combine this result with the recent $B^+ \to K^+ \tau^+\tau^-$~\cite{butokplustautauB1B2} measurement to determine $\mathcal{B}(B\to K \tau^+ \tau^-)<5.4 \times 10^{-4}$ at the 90\% confidence level, which represents the most stringent constraint on $b\to s \tau^+\tau^-$ transitions involving a pseudoscalar $K$ meson.

\clearpage
\clearpage

\clearpage
\clearpage

\section*{Acknowledgments}

\begingroup
\setlength{\emergencystretch}{2em} 
\sloppy  
This work, based on data collected using the Belle II detector, which was built and commissioned prior to March 2019,
and data collected using the Belle detector, which was operated until June 2010,
was supported by
Higher Education and Science Committee of the Republic of Armenia Grant No.~23LCG-1C011;
Australian Research Council and Research Grants
No.~DP200101792, 
No.~DP210101900, 
No.~DP210102831, 
No.~DE220100462, 
No.~LE210100098, 
and
No.~LE230100085; 
Austrian Federal Ministry of Education, Science and Research,
Austrian Science Fund (FWF) Grants
DOI:~10.55776/P34529,
DOI:~10.55776/J4731,
DOI:~10.55776/J4625,
DOI:~10.55776/M3153,
and
DOI:~10.55776/PAT1836324,
and
Horizon 2020 ERC Starting Grant No.~947006 ``InterLeptons'';
Natural Sciences and Engineering Research Council of Canada, Digital Research Alliance of Canada, and Canada Foundation for Innovation;
National Key R\&D Program of China under Contract No.~2024YFA1610503,
and
No.~2024YFA1610504,
National Natural Science Foundation of China and Research Grants
No.~11575017,
No.~11761141009,
No.~11705209,
No.~11975076,
No.~12135005,
No.~12150004,
No.~12161141008,
No.~12405099,
No.~12475093,
and
No.~12175041,
and Shandong Provincial Natural Science Foundation Project~ZR2022JQ02;
the Czech Science Foundation Grant No. 22-18469S,  Regional funds of EU/MEYS: OPJAK
FORTE CZ.02.01.01/00/22\_008/0004632 
and
Charles University Grant Agency project No. 246122;
European Research Council, Seventh Framework PIEF-GA-2013-622527,
Horizon 2020 ERC-Advanced Grants No.~267104 and No.~884719,
Horizon 2020 ERC-Consolidator Grant No.~819127,
Horizon 2020 Marie Sklodowska-Curie Grant Agreement No.~700525 ``NIOBE''
and
No.~101026516,
and
Horizon Europe Marie Sklodowska-Curie Staff Exchange project JENNIFER3 Grant Agreement No.~101183137 (European grants);
L’Institut National de Physique Nucl\'eaire et de Physique des
Particules (IN2P3) du CNRS under Project Identification No.
CNRS-IN2P3-14-PP-033
and L’Agence Nationale de la Recherche (ANR) under Grant No. ANR-23-CE31-0018 and ANR-25-CE31-1333 (France);
BMFTR, DFG, HGF, MPG, and AvH Foundation (Germany);
Department of Atomic Energy under Project Identification No.~RTI 4002,
Department of Science and Technology,
and
UPES SEED funding programs
No.~UPES/R\&D-SEED-INFRA/17052023/01 and
No.~UPES/R\&D-SOE/20062022/06 (India);
Israel Science Foundation Grant No.~2476/17,
U.S.-Israel Binational Science Foundation Grant No.~2016113, and
Israel Ministry of Science Grant No.~3-16543;
Istituto Nazionale di Fisica Nucleare and the Research Grants BELLE2,
and
the ICSC – Centro Nazionale di Ricerca in High Performance Computing, Big Data and Quantum Computing, funded by European Union – NextGenerationEU;
Japan Society for the Promotion of Science, Grant-in-Aid for Scientific Research Grants
No.~16H03993,
No.~16H06492,
No.~16K05323,
No.~17H01133,
No.~17H05405,
No.~18K03621,
No.~18H03710,
No.~18H05226,
No.~19H00682, 
No.~20H05850,
No.~20H05858,
No.~22H00144,
No.~22K14056,
No.~22K21347,
No.~23H05433,
No.~26220706,
No.~26400255,
and
No.~26H02056,
and
the Ministry of Education, Culture, Sports, Science, and Technology (MEXT) of Japan;  
National Research Foundation (NRF) of Korea Grants
No.~2021R1-F1A-1064008,
No.~2022R1-A2C-1003993,
No.~RS-2018-NR031074,
No.~RS-2021-NR060129,
No.~RS-2024-00354342,
No.~RS-2025-02219521,
No.~RS-2026-25471491,
No.~RS-2026-25480677,
and
No.~RS-2026-25486791,
Radiation Science Research Institute,
Foreign Large-Size Research Facility Application Supporting project,
the Global Science Experimental Data Hub Center, the Korea Institute of Science and
Technology Information (K26L1M2C3)
and
KREONET/GLORIAD;
Universiti Malaya RU grant, Akademi Sains Malaysia, and Ministry of Education Malaysia;
Frontiers of Science Program Contracts
No.~FOINS-296,
No.~CB-221329,
No.~CB-236394,
No.~CB-254409,
and
No.~CB-180023, and SEP-CINVESTAV Research Grant No.~237 (Mexico);
the Polish Ministry of Science and Higher Education and the National Science Center;
the Ministry of Science and Higher Education of the Russian Federation
and
the HSE University Basic Research Program, Moscow;
University of Tabuk Research Grants
No.~S-0256-1438 and No.~S-0280-1439 (Saudi Arabia);
Slovenian Research Agency and Research Grants
No.~J1-50010
and
No.~P1-0135;
Ikerbasque, Basque Foundation for Science,
State Agency for Research of the Spanish Ministry of Science and Innovation through Grant No. PID2022-136510NB-C33, Spain,
the Severo Ochoa project CEX2023-001292-S funded by MICIU/AEI, State Secretariat for
Telecommunications and Digital Infrastructure with reference
TSI-069100-2023-0012, State Agency for Research of the Spanish Ministry
of Science, Innovation and Universities through Grant No
PID2024-156645NB-C21;
the Swiss National Science Foundation;
The Knut and Alice Wallenberg Foundation (Sweden), Contracts No.~2021.0174, No.~2021.0299, and No.~2023.0315;
National Science and Technology Council,
and
Ministry of Education (Taiwan);
Thailand Center of Excellence in Physics;
TUBITAK ULAKBIM (Turkey);
National Research Foundation of Ukraine, Project No.~2020.02/0257,
and
Ministry of Education and Science of Ukraine;
the U.S. National Science Foundation and Research Grants
No.~PHY-1913789 
and
No.~PHY-2111604, 
and the U.S. Department of Energy and Research Awards
No.~DE-AC06-76RLO1830, 
No.~DE-SC0007983, 
No.~DE-SC0009824, 
No.~DE-SC0009973, 
No.~DE-SC0010007, 
No.~DE-SC0010073, 
No.~DE-SC0010118, 
No.~DE-SC0010504, 
No.~DE-SC0011784, 
No.~DE-SC0012704, 
No.~DE-SC0019230, 
No.~DE-SC0021616, 
No.~DE-SC0022350, 
No.~DE-SC0023470; 
and
the Vietnam Academy of Science and Technology (VAST) under Grant
No.~DL0000.05/26-27.

These acknowledgements are not to be interpreted as an endorsement of any statement made
by any of our institutes, funding agencies, governments, or their representatives.

We thank the SuperKEKB team for delivering high-luminosity collisions;
the KEK cryogenics group for the efficient operation of the detector solenoid magnet and IBBelle on site;
the KEK Computer Research Center for on-site computing support; the NII for SINET6 network support;
and the raw-data centres hosted by BNL, DESY, GridKa, IN2P3, INFN, 
PNNL/EMSL, 
and the University of Victoria.

\endgroup

\bibliographystyle{JHEP}
\bibliography{references}

@article{Bevan:2014iga,
      author         = "{Ed.~A.~J.~Bevan, B.~Golob, Th.~Mannel, S.~Prell, and B.~D.~Yabsley}",
      title          = "{The Physics of the \B Factories}",
      journal        = "Eur. Phys. J. C",
      volume         = "74",
      year           = "2014",
      pages          = "3026",
      doi            = "10.1140/epjc/s10052-014-3026-9",
      eprint         = "1406.6311",
      archivePrefix  = "arXiv",
      primaryClass   = "hep-ex",
      reportNumber   = "SLAC-PUB-15968, KEK-PREPRINT-2014-3,
                        FERMILAB-PUB-14-262-T",
      SLACcitation   = "%%CITATION = ARXIV:1406.6311;%%"
}

@article{Brodzicka:2012jm,
      author         = "Brodzicka, Jolanta and others",
      title          = "{Physics achievements from the Belle Experiment}",
      collaboration  = "Belle",
      journal        = "PTEP",
      volume         = "2012",
      year           = "2012",
      pages          = "04D001",
      doi            = "10.1093/ptep/pts072",
      eprint         = "1212.5342",
      archivePrefix  = "arXiv",
      primaryClass   = "hep-ex",
      reportNumber   = "KEK-REPORT-2012-5",
      SLACcitation   = "%%CITATION = ARXIV:1212.5342;%%"
}

@article{Abe:2010gxa,
      author         = "Abe, T.",
      title          = "{Belle II technical design report}",
      collaboration  = "Belle II ",
      year           = "2010",
      eprint         = "1011.0352",
      archivePrefix  = "arXiv",
      primaryClass   = "physics.ins-det",
      reportNumber   = "KEK-REPORT-2010-1",
      SLACcitation   = "%%CITATION = ARXIV:1011.0352;%%",
}

@article{Akai:2018mbz,
      author         = "Akai, Kazunori and Furukawa, Kazuro and Koiso, Haruyo",
      title          = "{SuperKEKB collider}",
      journal        = "Nucl. Instrum. Meth. A",
      volume         = "907",
      year           = "2018",
      pages          = "188",
      doi            = "10.1016/j.nima.2018.08.017",
      eprint         = "1809.01958",
      archivePrefix  = "arXiv",
      primaryClass   = "physics.acc-ph",
      SLACcitation   = "%%CITATION = ARXIV:1809.01958;%%"
}

@article{Ohnishi:2013fma,
    author = "Ohnishi, Yukiyoshi and others",
    title = "{Accelerator design at SuperKEKB}",
    doi = "10.1093/ptep/pts083",
    journal = "PTEP",
    volume = "2013",
    pages = "03A011",
    year = "2013"
}

@article{Belle-IISVD:2022upf,
    author = "Adamczyk, K. and others",
    collaboration = "Belle II SVD",
    title = "{The design, construction, operation and performance of the Belle~II silicon vertex detector}",
    eprint = "2201.09824",
    archivePrefix = "arXiv",
    primaryClass = "physics.ins-det",
    doi = "10.1088/1748-0221/17/11/P11042",
    journal = "JINST",
    volume = "17",
    number = "11",
    pages = "P11042",
    year = "2022"
}

@article{Abashian2002117,
    eprint = "",
    title = "{The Belle Detector}",
    journal = "Nucl. Instrum. Meth. A",
    volume = {479},
    number = {1},
    pages = {117-232},
    year = {2002},
    doi = {https://doi.org/10.1016/S0168-9002(01)02013-7},
    url = {https://www.sciencedirect.com/science/article/pii/S0168900201020137},
    author = "Abashian, A. and others"
}

@article{Kurokawa:2001nw,
    author = "Kurokawa, S. and Kikutani, Eiji",
    title = "{Overview of the KEKB accelerators}",
    reportNumber = "KEK-PREPRINT-2001-157A",
    doi = "10.1016/S0168-9002(02)01771-0",
    journal = "Nucl. Instrum. Meth. A",
    volume = "499",
    pages = "1--7",
    year = "2003"
}

@article{Abe:2013kxa,
    author = "Abe, Tetsuo and others",
    title = "{Achievements of KEKB}",
    doi = "10.1093/ptep/pts102",
    journal = "PTEP",
    volume = "2013",
    pages = "03A001",
    year = "2013"
}

@article{Lange:2001uf,
      author         = "Lange, D. J.",
      title          = "{The EvtGen particle decay simulation package}",
      booktitle      = "{Proceedings, 7th International Conference on \B physics
                        at hadron machines (BEAUTY 2000): Maagan, Israel,
                        September 13-18, 2000}",
      journal        = "Nucl. Instrum. Meth. A",
      volume         = "462",
      year           = "2001",
      pages          = "152",
      doi            = "10.1016/S0168-9002(01)00089-4",
      SLACcitation   = "%%CITATION = NUIMA,A462,152;%%"
}

@article{KATAYAMA199822,
title = "{Belle computing model}",
journal = {Comput. Phys. Commun.},
volume = {110},
number = {1},
pages = {22-25},
year = {1998},
issn = {0010-4655},
doi = {https://doi.org/10.1016/S0010-4655(97)00148-3},
url = {https://www.sciencedirect.com/science/article/pii/S0010465597001483},
author = {Nobu Katayama and Ryosuke Itoh and Atsushi Manabe and Takashi Sasaki}
}

@article{Brun:1994aa,
    author = "Brun, Ren{\'e} and Bruyant, F. and Carminati, Federico and Giani, Simone and Maire, M. and McPherson, A. and Patrick, G. and Urban, L.",
    title = "{GEANT Detector Description and Simulation Tool}",
    journal = "CERN-W5013",
    doi = "10.17181/CERN.MUHF.DMJ1",
    month = "10",
    year = "1994",
    pages = "{\unskip}",
}

@article{Sjostrand:2014zea,
      author         = {Sj\"{o}strand, Torbj\"{o}rn and Ask, Stefan and Christiansen,
                        Jesper R. and Corke, Richard and Desai, Nishita and Ilten,
                        Philip and Mrenna, Stephen and Prestel, Stefan and
                        Rasmussen, Christine O. and Skands, Peter Z.},
      title          = "{An Introduction to PYTHIA 8.2}",
      journal        = "Comput. Phys. Commun.",
      volume         = "191",
      year           = "2015",
      pages          = "159-177",
      doi            = "10.1016/j.cpc.2015.01.024",
      eprint         = "1410.3012",
      archivePrefix  = "arXiv",
      primaryClass   = "hep-ph",
      reportNumber   = "LU-TP-14-36, MCNET-14-22, CERN-PH-TH-2014-190,
                        FERMILAB-PUB-14-316-CD, DESY-14-178, SLAC-PUB-16122",
      SLACcitation   = "%%CITATION = ARXIV:1410.3012;%%"
}

@article{uffi,
doi = {10.1088/1126-6708/2006/05/026},
url = {https://dx.doi.org/10.1088/1126-6708/2006/05/026},
year = {2006},
month = {may},
publisher = {},
volume = {05},
number = {05},
pages = {026},
author = {Torbjörn Sjöstrand and Stephen Mrenna and Peter Skands},
title = "{PYTHIA 6.4 physics and manual}",
journal = {JHEP}
}

@article{Jadach:1999vf,
      author         = "Jadach, S. and Ward, B. F. L. and W\c{a}s, Z.",
      title          = "{The precision Monte Carlo event generator KK for two-fermion
                        final states in $e^+e^-$ collisions}",
      journal        = "Comput. Phys. Commun.",
      volume         = "130",
      year           = "2000",
      pages          = "260",
      doi            = "10.1016/S0010-4655(00)00048-5",
      eprint         = "hep-ph/9912214",
      archivePrefix  = "arXiv",
      primaryClass   = "hep-ph",
      reportNumber   = "DESY-99-106, CERN-TH-99-235, UTHEP-99-08-01",
      SLACcitation   = "%%CITATION = HEP-PH/9912214;%%"
}

@article{Barberio:1993qi,
    author       = "Barberio, Elisabetta and W\c{a}s, Zbigniew",
    title        = "{PHOTOS: A Universal Monte Carlo for QED radiative 
                     corrections. Version 2.0}",
    reportNumber = "CERN-TH-7033-93",
    doi          = "10.1016/0010-4655(94)90074-4",
    journal      = "Comput. Phys. Commun.",
    volume       = "79",
    pages        = "291",
    year         = "1994"
}

@article{Agostinelli:2002hh,
      author         = "Agostinelli, S. and others",
      title          = "{GEANT4: A simulation toolkit}",
      collaboration  = "GEANT4 ",
      journal        = "Nucl. Instrum. Meth. A",
      volume         = "506",
      pages          = "250-303",
      doi            = "10.1016/S0168-9002(03)01368-8",
      year           = "2003",
      reportNumber   = "SLAC-PUB-9350, FERMILAB-PUB-03-339",
      SLACcitation   = "%%CITATION = NUIMA,A506,250;%%",
}

@article{Kuhr:2018lps,
      author         = "Kuhr, T. and Pulvermacher, C. and Ritter, M. and Hauth,
                        T. and Braun, N.",
      title          = "{The Belle II Core Software}",
      collaboration  = "Belle II Framework Software Group",
      journal        = "Comput. Softw. Big Sci.",
      volume         = "3",
      year           = "2019",
      number         = "1",
      pages          = "1",
      doi            = "10.1007/s41781-018-0017-9",
      eprint         = "1809.04299",
      archivePrefix  = "arXiv",
      primaryClass   = "physics.comp-ph",
      SLACcitation   = "%%CITATION = ARXIV:1809.04299;%%"
}

@article{keck,
      author         = "Keck, T. and others",
      title          = "{The Full Event Interpretation}",
      journal        = "Comput. Softw. Big Sci.",
      volume         = "3",
      year           = "2019",
      number         = "1",
      pages          = "6",
      doi            = "10.1007/s41781-019-0021-8",
      eprint         = "1807.08680",
      archivePrefix  = "arXiv",
      primaryClass   = "hep-ex",
      reportNumber   = "DESY-19-048",
      SLACcitation   = "%%CITATION = ARXIV:1807.08680;%%"
}

@article{ParticleDataGroup:2024cfk,
    author = "Navas, S. and others",
    collaboration = "Particle Data Group",
    title = "{Review of particle physics}",
    doi = "10.1103/PhysRevD.110.030001",
    journal = "Phys. Rev. D",
    volume = "110",
    number = "3",
    pages = "030001",
    year = "2024"
}

@article{Atmacan:2025jmh,
    author = "Atmacan, H. and others",
    title = "{The imaging Time-of-Propagation detector at Belle II}",
    eprint = "2504.19090",
    archivePrefix = "arXiv",
    primaryClass = "hep-ex",
    reportNumber = "UCHEP-25-01, University of Cincinnati preprint UCHEP-25-01",
    doi = "10.1016/j.nima.2025.170627",
    journal = "Nucl. Instrum. Meth. A",
    volume = "1080",
    pages = "170627",
    year = "2025"
}

@article{Parrott,
  title = "{Standard Model predictions for $B\ensuremath{\rightarrow}K{\ensuremath{\ell}}^{+}{\ensuremath{\ell}}^{\ensuremath{-}}$, $B\ensuremath{\rightarrow}K{\ensuremath{\ell}}_{1}^{\ensuremath{-}}{\ensuremath{\ell}}_{2}^{+}$ and $B\ensuremath{\rightarrow}K\ensuremath{\nu}\overline{\ensuremath{\nu}}$ using form factors from ${N}_{f}=2+1+1$ lattice QCD}",
  author = {Parrott, W. G. and Bouchard, C. and Davies, C. T. H.},
  collaboration = {HPQCD},
  journal = {Phys. Rev. D},
  volume = {107},
  issue = {1},
  pages = {014511},
  numpages = {28},
  year = {2023},
  publisher = {American Physical Society},
  doi = {10.1103/PhysRevD.107.014511},
  url = {https://link.aps.org/doi/10.1103/PhysRevD.107.014511},
  note={[Erratum: \href{https://doi.org/10.1103/PhysRevD.107.119903}{\textit{Phys. Rev. D} \textbf{107} (2023) 119903}]},
}

@article{btoknunu,
  title = "{Evidence for ${B}^{+}\ensuremath{\rightarrow}{K}^{+}\ensuremath{\nu}\overline{\ensuremath{\nu}}$ decays}",
  author = {Adachi, I. and Adamczyk, K. and Aggarwal, L. and Ahmed, H. and Aihara, H. and Akopov, N. and Aloisio, A. and Anh Ky, N. and others},
  collaboration = {Belle II },
  journal = {Phys. Rev. D},
  volume = {109},
  issue = {11},
  pages = {112006},
  numpages = {29},
  year = {2024},
  eprint = "2311.14647",
  publisher = {American Physical Society},
  doi = {10.1103/PhysRevD.109.112006},
  url = {https://link.aps.org/doi/10.1103/PhysRevD.109.112006}
}

@article{bdtokstartautau,
  title = "{Search for the decay ${B}^{0}\ensuremath{\rightarrow}{K}^{*0}{\ensuremath{\tau}}^{+}{\ensuremath{\tau}}^{\ensuremath{-}}$ at the Belle Experiment}",
  author = {Dong, T. V. and Luo, T. and Adachi, I. and Aihara, H. and Asner, D. M. and others},
  collaboration = {Belle },
  journal = {Phys. Rev. D},
  volume = {108},
  issue = {1},
  pages = {L011102},
  numpages = {9},
  year = {2023},
  eprint = "2110.03871",
  publisher = {American Physical Society},
  doi = {10.1103/PhysRevD.108.L011102},
  url = {https://link.aps.org/doi/10.1103/PhysRevD.108.L011102}
}

@article{butokplustautau,
  title = "{Search for ${B}^{+}\ensuremath{\rightarrow}{K}^{+}{\ensuremath{\tau}}^{+}{\ensuremath{\tau}}^{\ensuremath{-}}$ at the BaBar Experiment}",
  author = {Lees, J. P. and Poireau, V. and Tisserand, V. and Grauges, E. and Palano, A. and others},
  collaboration = {BaBar },
  journal = {Phys. Rev. Lett.},
  volume = {118},
  issue = {3},
  pages = {031802},
  numpages = {8},
  year = {2017},
  eprint = "1605.09637",
  publisher = {American Physical Society},
  doi = {10.1103/PhysRevLett.118.031802},
  url = {https://link.aps.org/doi/10.1103/PhysRevLett.118.031802}
}

@article{butokplustautauB1B2,
  title = {Search for the decay ${B}^+ \rightarrow {K}^+\tau^+\tau^-$ using data from the {Belle} and {Belle~II} {E}xperiments},
  author = {Abumusabh, M. and Adachi, I. and Adamczyk, K. and Aggarwal, A. and Aggarwal, L. and Ahmed, H. and Ahn, Y. and Aihara, H. and Akopov, N. and Alghamdi, S. and Alhakami, M. and Aloisio, A. and Althubiti, N. and Amos, K. and Angelsmark, M. and Anh Ky, N. and Antonioli, C. and Asner, D. M. and Atmacan, H. and Aushev, T. and Ayad, R. and Babu, V. and Bae, H. and Baghel, N. K. and Bahinipati, S. and Bambade, P. and Banerjee, Sw. and Barrett, M. and Bartl, M. and Baudot, J. and Baur, A. and Beaubien, A. and Becherer, F. and Becker, J. and Bennett, J. V. and Bernlochner, F. U. and Bertacchi, V. and Bertemes, M. and Bertholet, E. and Bessner, M. and Bettarini, S. and Bhardwaj, V. and Bianchi, F. and Bilka, T. and Biswas, D. and Bobrov, A. and Bodrov, D. and Bondar, A. and Bonvicini, G. and Borah, J. and Boschetti, A. and Bozek, A. and Bra{\ifmmode \check{c}\else \v{c}\fi{}}ko, M. and Branchini, P. and Briere, R. A. and Browder, T. E. and Budano, A. and Bussino, S. and Campagna, Q. and Campajola, M. and Cao, L. and Casarosa, G. and Cecchi, C. and Chang, M.-C. and Chang, P. and Cheema, P. and Chen, L. and Cheon, B. G. and Cheshta, C. and Chetri, H. and Chilikin, K. and Chirapatpimol, K. and Cho, H.-E. and Cho, K. and Cho, S.-J. and Choi, S.-K. and Choudhury, S. and Chutia, S. and Cochran, J. and Colorado-Caicedo, J. A. and Consigny, I. and Corona, L. and Cui, J. X. and De La Cruz-Burelo, E. and De La Motte, S. A. and de Marino, G. and De Nardo, G. and De Pietro, G. and de Sangro, R. and Destefanis, M. and Dey, S. and Dhayal, R. and Di Canto, A. and Dingfelder, J. and Dole{\ifmmode \check{z}\else \v{z}\fi{}}al, Z. and Dom{\'{\i}}nguez Jim{\'e}nez, I. and Dong, T. V. and Dong, X. and Dorigo, M. and Dugic, K. and Dujany, G. and Ecker, P. and Eppelt, J. and Farkas, R. and Ferber, T. and Fillinger, T. and Finck, C. and Finocchiaro, G. and Forti, F. and Frey, A. and Fulsom, B. G. and Gabrielli, A. and Gale, A. and Ganiev, E. and Garcia-Hernandez, M. and Garg, R. and G{\"a}rtner, L. and Gaudino, G. and Gaur, V. and Gautam, V. and Gaz, A. and Gellrich, A. and Ghevondyan, G. and Ghosh, D. and Giordano, R. and Giri, A. and Gironella Gironell, P. and Glazov, A. and Gobbo, B. and Godang, R. and Gogota, O. and Goldenzweig, P. and Gradl, W. and Graziani, E. and Greenwald, D. and Guan, Y. and Gudkova, K. and Haide, I. and Haigh, H. and Han, Y. and Hayashii, H. and Hazra, S. and Hearty, C. and Hedges, M. T. and Heidelbach, A. and Heine, G. and Heredia de la Cruz, I. and Hern{\'a}ndez Villanueva, M. and Higuchi, T. and Hoek, M. and Hohmann, M. and Hoppe, R. and Horak, P. and Hou, X. T. and Hsu, C.-L. and Humair, T. and Iijima, T. and Inami, K. and Inguglia, G. and Ipsita, N. and Ishikawa, A. and Itoh, R. and Iwasaki, M. and Jackson, P. and Jacobi, D. and Jacobs, W. W. and Jang, E.-J. and Ji, Q. P. and Jia, S. and Jin, Y. and Johnson, A. and Kandra, J. and Kang, K. H. and Kang, S. and Karyan, G. and Keil, F. and Ketter, C. and Kiesling, C. and Kim, C. and Kim, D. Y. and Kim, H. and Kim, J.-Y. and Kim, K.-H. and Kindo, H. and Kinoshita, K. and Kody{\ifmmode \check{s}\else \v{s}\fi{}}, P. and Koga, T. and Kohani, S. and Korobov, A. and Korpar, S. and Kovalenko, E. and Kowalewski, R. and Kri{\ifmmode \check{z}\else \v{z}\fi{}}an, P. and Krokovny, P. and Kuhr, T. and Kulii, Y. and Kumar, D. and Kumar, J. and Kumar, R. and Kumara, K. and Kunigo, T. and Kuzmin, A. and Kwon, Y.-J. and Lacaprara, S. and Lam, T. and Lanceri, L. and Lange, J. S. and Lau, T. S. and Laurenza, M. and Leboucher, R. and Le Diberder, F. R. and Lee, H. and Lee, M. J. and Lemettais, C. and Leo, P. and Lewis, P. M. and Li, C. and Li, H.-J. and Li, L. K. and Li, Q. M. and Li, W. Z. and Li, Y. and Li, Y. B. and Liao, Y. P. and Libby, J. and Lin, J. and Lin, S. and Liptak, Z. and Lisovskyi, V. and Liu, M. H. and Liu, Q. Y. and Liu, Y. and Liu, Z. Q. and Liventsev, D. and Longo, S. and Lozar, A. and Lueck, T. and Lyu, C. and Ma, J. L. and Ma, Y. and Maggiora, M. and Maharana, S. P. and Maiti, R. and Mancinelli, G. and Manfredi, R. and Manoni, E. and Mantovano, M. and Marcantonio, D. and Marcello, S. and Marfoli, M. and Marinas, C. and Martellini, C. and Martens, A. and Martinov, T. and Massaccesi, L. and Masuda, M. and Matvienko, D. and Maurya, S. K. and Maushart, M. and McKenna, J. A. and Mediankin Gruberov{\'a}, Z. and Mehta, R. and Meier, F. and Meleshko, D. and Merola, M. and Miller, C. and Mirra, M. and Miyabayashi, K. and Miyake, H. and Mizuk, R. and Mohanty, G. B. and Moneta, S. and Moreira de Carvalho, A. L. and Moser, H.-G. and Muller, Th. and Mussa, R. and Nakamura, I. and Nakao, M. and Nakazawa, Y. and Naruki, M. and Natkaniec, Z. and Natochii, A. and Nayak, M. and Neu, M. and Niiyama, M. and Nishida, S. and Nomaru, R. and Ogawa, S. and Okubo, R. and Ono, H. and Onuki, Y. and Otani, F. and Pakhlov, P. and Pakhlova, G. and Panta, A. and Pardi, S. and Parham, K. and Park, J. and Park, K. and Park, S.-H. and Passeri, A. and Patra, S. and Paul, S. and Pedlar, T. K. and Pestotnik, R. and Piccolo, M. and Piilonen, L. E. and Podesta-Lerma, P. L. M. and Podobnik, T. and Prakash, A. and Praz, C. and Prell, S. and Prencipe, E. and Prim, M. T. and Privalov, S. and Prudiiev, I. and Purwar, H. and Rados, P. and Raiz, S. and Ravindran, K. and Rehman, J. U. and Reif, M. and Reiter, S. and Reuter, L. and Ricalde Herrmann, D. and Ripp-Baudot, I. and Rizzo, G. and Robertson, S. H. and Roney, J. M. and Rostomyan, A. and Rout, N. and Saha, S. and Salutari, L. and Sanders, D. A. and Sandilya, S. and Santelj, L. and Santos, C. and Savinov, V. and Scavino, B. and Schmitt, C. and Schneider, S. and Schnell, G. and Schoenning, K. and Schwanda, C. and Schwartz, A. J. and Seino, Y. and Senyo, K. and Serrano, J. and Sevior, M. E. and Sfienti, C. and Shan, W. and Sharma, G. and Shen, C. P. and Shi, X. D. and Shillington, T. and Shimasaki, T. and Shiu, J.-G. and Shtol, D. and Shwartz, B. and Sibidanov, A. and Simon, F. and Skorupa, J. and Sobie, R. J. and Sobotzik, M. and Soffer, A. and Sokolov, A. and Solovieva, E. and Song, W. and Spataro, S. and {\ifmmode \check{S}\else \v{S}\fi{}}penko, K. and Spruck, B. and Stari{\ifmmode \check{c}\else \v{c}\fi{}}, M. and Stavroulakis, P. and Stefkova, S. and Stroili, R. and Strube, J. and Sumihama, M. and Suwonjandee, N. and Takahashi, M. and Takizawa, M. and Tang, S. S. and Tanida, K. and Tenchini, F. and Testa, F. and Thaller, A. and Tien Manh, T. and Tittel, O. and Tiwary, R. and Tonelli, D. and Torassa, E. and Trabelsi, K. and Trantou, F. F. and Tsaklidis, I. and Ueda, I. and Uglov, T. and Unger, K. and Unno, Y. and Uno, K. and Uno, S. and Urquijo, P. and Ushiroda, Y. and Vahsen, S. E. and van Tonder, R. and Varvell, K. E. and Veronesi, M. and Vismaya, V. S. and Vitale, L. and Vobbilisetti, V. and Volpe, R. and Wakai, M. and Wallner, S. and Wang, M.-Z. and Warburton, A. and Watanabe, M. and Watanuki, S. and Wessel, C. and Won, E. and Xu, X. P. and Yabsley, B. D. and Yan, W. and Yan, W. and Yelton, J. and Yi, K. and Yin, J. H. and Yoshihara, K. and Yuan, J. and Yusa, Y. and Zani, L. and Zeng, F. and Zeyrek, M. and Zhang, B. and Zhilich, V. and Zhou, J. S. and Zhou, Q. D. and Zhu, L. and {\ifmmode \check{Z}\else \v{Z}\fi{}}leb{\ifmmode \check{c}\else \v{c}\fi{}}{\'{\i}}k, R.},
  collaboration = {Belle~II, Belle},
  journal = {Phys. Rev. Lett.},
  volume = {137},
  issue = {5},
  pages = {051805},
  numpages = {8},
  year = {2026},
  month = {Jul},
  publisher = {American Physical Society},
  doi = {10.1103/12f4-z51q},
  url = {https://link.aps.org/doi/10.1103/12f4-z51q},
  eprint = "2603.24437",
  archivePrefix = "arXiv",

}

@article{Capdevila,
  title = "{Searching for New Physics with $b\ensuremath{\rightarrow}s{\ensuremath{\tau}}^{+}{\ensuremath{\tau}}^{\ensuremath{-}}$ Processes}",
  author = {Capdevila, Bernat and Crivellin, Andreas and Descotes-Genon, S\'ebastien and Hofer, Lars and Matias, Joaquim},
  journal = {Phys. Rev. Lett.},
  volume = {120},
  issue = {18},
  pages = {181802},
  numpages = {7},
  year = {2018},
  eprint = "1712.01919",
  publisher = {American Physical Society},
  doi = {10.1103/PhysRevLett.120.181802},
  url = {https://link.aps.org/doi/10.1103/PhysRevLett.120.181802}
}

@article{Allwicher,
	author = {L. Allwicher and D. Be{\v c}irevi{\'c} and G. Piazza and S. Rosauro-Alcaraz and O. Sumensari},
	doi = {https://doi.org/10.1016/j.physletb.2023.138411},
	issn = {0370-2693},
	journal = {Phys. Lett. B},
	pages = {138411},
	title = "{Understanding the first measurement of $\mathcal{B}({B}\to {K} \nu\bar{\nu})$}",
	url = {https://www.sciencedirect.com/science/article/pii/S037026932300744X},
	volume = {848},
	year = {2024},
    eprint = "2309.02246"}

@article{Bause,
    author = "Bause, Rigo and Gisbert, Hector and Hiller, Gudrun",
    title = "{Implications of an enhanced B{\textrightarrow}K{\ensuremath{\nu}}{\ensuremath{\bar{\nu}}} branching ratio}",
    eprint = "2309.00075",
    archivePrefix = "arXiv",
    primaryClass = "hep-ph",
    doi = "10.1103/PhysRevD.109.015006",
    journal = "Phys. Rev. D",
    volume = "109",
    number = "1",
    pages = "015006",
    year = "2024"
}

@article{fbdt,
	author = {Keck, Thomas},
	date = {2017/09/29},
	doi = {10.1007/s41781-017-0002-8},
	id = {Keck2017},
	isbn = {2510-2044},
	journal = {Comput. Softw. Big Sci.},
	number = {1},
	pages = {2},
	title = "{FastBDT: A Speed-Optimized Multivariate Classification Algorithm for the Belle II Experiment}",
	url = {https://doi.org/10.1007/s41781-017-0002-8},
	volume = {1},
	year = {2017}}

@article{mutransf,
	author = {Angus, John E.},
	doi = {10.1137/1036146},
	journal = {SIAM Review},
	number = {4},
	pages = {652-654},
	title = "{The Probability Integral Transform and Related Results}",
	volume = {36},
	year = {1994}
    }

@article{caby,
	author = {Cranmer, Kyle and Held, Alexander},
	doi = {10.1051/epjconf/202125103067},
	journal = {EPJ Web Conf.},
	pages = {03067},
	title = {Building and steering binned template fits with cabinetry},
	url = {https://doi.org/10.1051/epjconf/202125103067},
	volume = 251,
	year = 2021}

@article{pyhf_joss,
  doi = {10.21105/joss.02823},
  url = {https://doi.org/10.21105/joss.02823},
  year = {2021},
  publisher = {The Open Journal},
  volume = {6},
  number = {58},
  pages = {2823},
  author = {Lukas Heinrich and Matthew Feickert and Giordon Stark and Kyle Cranmer},
  title = "{pyhf: pure-Python implementation of HistFactory statistical models}",
  journal = {JOSS}
}

@article{f00hflav,
  author = "Banerjee, Sw. and others",
  collaboration = "Heavy Flavor Averaging Group (HFLAV)",
  title = "{Averages of $b$-hadron, $c$-hadron, and $\tau$-lepton properties as of 2023}",
  eprint = "2411.18639",
  archivePrefix = "arXiv",
  primaryClass = "hep-ex",
  doi = "10.1103/x87q-tld5",
  journal = "Phys. Rev. D",
  volume = "113",
  number = "1",
  pages = "012008",
  year = "2026"
}

@article{btosllball,
  title = "{Comparative study of the decays ${B} \to {(K,K}^{*}{)\ell}^{+}{\ell}^{\ensuremath{-}}$ in the {S}tandard {M}odel and {S}upersymmetric theories}",
  author = {Ali, A. and Ball, Patricia and Handoko, L. T. and Hiller, G.},
  journal = {Phys. Rev. D},
  volume = {61},
  issue = {7},
  pages = {074024},
  numpages = {19},
  year = {2000},
  eprint = "hep-ph/9910221",
  publisher = {American Physical Society},
  doi = {10.1103/PhysRevD.61.074024},
  url = {https://link.aps.org/doi/10.1103/PhysRevD.61.074024}
}

@article{Straub:2018kue,
    author = "Straub, David M.",
    title = "{flavio: a Python package for flavour and precision phenomenology in the Standard Model and beyond}",
    eprint = "1810.08132",
    archivePrefix = "arXiv",
    primaryClass = "hep-ph",
    month = "10",
    year = "2018"
}

@article{gBKS,
  title = "{Measurement of ${D}^{+}\ensuremath{\rightarrow}{K}_{S}^{0}{K}^{+}$ and ${D}_{s}^{+}\ensuremath{\rightarrow}{K}_{S}^{0}{\ensuremath{\pi}}^{+}$ branching ratios}",
  author = {Won, E. and Ko, B. R. and Aihara, H. and Arinstein, K. and Aulchenko, V. and Aushev, T. and Bakich, A. M. and Balagura, V. and Barberio, E. and Bay, A. and Belous, K. and Bhardwaj, V. and Bischofberger, M. and Bondar, A. and Bozek, A. and Bra\ifmmode \check{c}\else \v{c}\fi{}ko, M. and Browder, T. E. and Chang, P. and Chen, A. and Chen, P. and Cheon, B. G. and Chiang, C.-C. and Cho, I.-S. and Choi, Y. and Dalseno, J. and Das, A. and Eidelman, S. and Epifanov, D. and Esen, S. and Gabyshev, N. and Garmash, A. and Golob, B. and Ha, H. and Haba, J. and Han, B.-Y. and Hasegawa, Y. and Hayasaka, K. and Hayashii, H. and Hoshi, Y. and Hou, W.-S. and Hsiung, Y. B. and Hyun, H. J. and Iijima, T. and Inami, K. and Itoh, R. and Iwasaki, M. and Iwasaki, Y. and Joshi, N. J. and Julius, T. and Kang, J. H. and Katayama, N. and Kawasaki, T. and Kiesling, C. and Kim, H. J. and Kim, H. O. and Kim, J. H. and Kim, S. K. and Kim, Y. I. and Kim, Y. J. and Korpar, S. and Krokovny, P. and Kumita, T. and Kuzmin, A. and Kwon, Y.-J. and Kyeong, S.-H. and Lange, J. S. and Lee, M. J. and Lee, S.-H. and Li, J. and Liu, C. and Liu, Y. and Liventsev, D. and Louvot, R. and Mandl, F. and McOnie, S. and Miyata, H. and Miyazaki, Y. and Mori, T. and Nakano, E. and Nakao, M. and Nakazawa, H. and Natkaniec, Z. and Nishida, S. and Nitoh, O. and Ohshima, T. and Okuno, S. and Pakhlov, P. and Pakhlova, G. and Palka, H. and Park, C. W. and Park, H. and Park, H. K. and Park, K. S. and Peak, L. S. and Pestotnik, R. and Petri\ifmmode \check{c}\else \v{c}\fi{}, M. and Piilonen, L. E. and Poluektov, A. and Ryu, S. and Sahoo, H. and Sakai, Y. and Schneider, O. and Schwanda, C. and Sevior, M. E. and Shapkin, M. and Shebalin, V. and Shiu, J.-G. and Shwartz, B. and Smerkol, P. and Sokolov, A. and Solovieva, E. and Stani\ifmmode \check{c}\else \v{c}\fi{}, S. and Stari\ifmmode \check{c}\else \v{c}\fi{}, M. and Sumiyoshi, T. and Taylor, G. N. and Teramoto, Y. and Trabelsi, K. and Uehara, S. and Unno, Y. and Uno, S. and Urquijo, P. and Usov, Y. and Varner, G. and Varvell, K. E. and Vervink, K. and Vinokurova, A. and Wang, C. H. and Wang, P. and Watanabe, Y. and Wedd, R. and Yabsley, B. D. and Yamashita, Y. and Yamauchi, M. and Zhang, C. C. and Zhang, Z. P. and Zhilich, V. and Zhulanov, V. and Zivko, T. and Zupanc, A. and Zyukova, O.},
  collaboration = {Belle},
  journal = {Phys. Rev. D},
  volume = {80},
  issue = {11},
  pages = {111101},
  numpages = {5},
  year = {2009},
  month = {Dec},
  eprint = "0910.3052",
  publisher = {American Physical Society},
  doi = {10.1103/PhysRevD.80.111101},
  url = {https://link.aps.org/doi/10.1103/PhysRevD.80.111101}
}

@article{bdtokstartautauB2,
    author = "Adachi, I. and others",
    collaboration = "Belle~II",
    title = "{Search for $B^0 \to K^{*0}\tau^+\tau^-$ Decays at the Belle II Experiment}",
    eprint = "2504.10042",
    archivePrefix = "arXiv",
    primaryClass = "hep-ex",
    reportNumber = "Belle II Preprint 2025-010; KEK Preprint 2025-8",
    doi = "10.1103/v1q3-9dy8",
    journal = "Phys. Rev. Lett.",
    volume = "135",
    number = "15",
    pages = "151801",
    year = "2025"
}

@article{Gelb:2018agf,
    author = "Gelb, Moritz and others",
    title = "{B2BII: Data Conversion from Belle to Belle II}",
    eprint = "1810.00019",
    archivePrefix = "arXiv",
    primaryClass = "hep-ex",
    doi = "10.1007/s41781-018-0016-x",
    journal = "Comput. Softw. Big Sci.",
    volume = "2",
    number = "1",
    pages = "9",
    year = "2018"
}

@article{LIPTAK2022167168,
title = "{Measurements of beam backgrounds in SuperKEKB Phase 2}",
journal = {Nucl. Instrum. Meth. A},
volume = {1040},
pages = {167168},
year = {2022},
issn = {0168-9002},
doi = {https://doi.org/10.1016/j.nima.2022.167168},
url = {https://www.sciencedirect.com/science/article/pii/S0168900222005149},
author = {Z. Liptak and A. Paladino and L. Santelj and J. Schueler and S. Stefkova and H. Tanigawa and N. Tsuzuki and A. Aloisio and P. Ahlburg and P. Bambade and G. Bassi and M. Barrett and J. Baudot and T.E. Browder and G. Casarosa and G. Cautero and D. Cinabro and G. Claus and D. Cuesta and F. {Di Capua} and S. {Di Carlo} and J. Flanagan and A. Frey and B.G. Fulsom and Y. Funakoshi and M. Gabriel and R. Giordano and D. Giuressi and M. Goffe and K. Hara and O. Hartbrich and M.T. Hedges and D. Heuchel and N. Iida and T. Ishibashi and K. Jaaskelainen and D. Jehanno and S. {de Jong} and T. Kraetzschmar and C. {La Licata} and L. Lanceri and P. Leitl and P.M. Lewis and C. Marinas and C. Miller and H. Moser and K.R. Nakamura and H. Nakayama and C. Niebuhr and Y. Onuki and C. Pang and B. Paschen and I. Ripp-Baudot and G. Rizzo and J.M. Roney and H. Schreeck and B. Schwenker and F. Simon and M. Specht and B. Spruck and Y. Soloviev and M. Szelezniak and S. Tanaka and S. Terui and G. Tortone and T. Tsuboyama and Y. Uematsu and S.E. Vahsen and L. Vitale and H. Windel}
}

@article{anom1,
  title = "{Global fit to $b\ensuremath{\rightarrow}c\ensuremath{\tau}\ensuremath{\nu}$ anomalies as of Spring 2024}",
  author = {Iguro, Syuhei and Kitahara, Teppei and Watanabe, Ryoutaro},
  journal = {Phys. Rev. D},
  volume = {110},
  issue = {7},
  pages = {075005},
  numpages = {25},
  year = {2024},
  month = {Oct},
  eprint = "2405.06062",
  publisher = {American Physical Society},
  doi = {10.1103/PhysRevD.110.075005},
  url = {https://link.aps.org/doi/10.1103/PhysRevD.110.075005}
}

@article{anom2,
	author = {Athron, Peter and Martinez, R. and Sierra, Cristian},
	date = {2024/02/16},
	doi = {10.1007/JHEP02(2024)121},
	id = {Athron2024},
	isbn = {1029-8479},
	journal = {JHEP},
	number = {2},
	pages = {121},
    eprint = "2308.13426",
	title = "{${B}$ meson anomalies and large ${B}^+ \to {K}^+\nu\bar{\nu}$ in non-universal ${U}(1)^\prime$ models}",
	url = {https://doi.org/10.1007/JHEP02(2024)121},
	volume = {02},
	year = {2024}}

@article{CLs1, 
doi = {10.1088/0954-3899/28/10/313},
url = {https://doi.org/10.1088/0954-3899/28/10/313},
year = {2002},
month = {sep},
publisher = {},
volume = {28},
number = {10},
pages = {2693},
author = {A L Read},
title = {Presentation of search results: the {CLs} technique},
journal = {J. Phys. G},

}

@article{CLs2,
	author = {Cowan, Glen and Cranmer, Kyle and Gross, Eilam and Vitells, Ofer},
	date = {2011/02/09},
	doi = {10.1140/epjc/s10052-011-1554-0},
	id = {Cowan2011},
	isbn = {1434-6052},
	journal = {Eur. Phys. J. C},
	number = {2},
	pages = {1554},
	title = "{Asymptotic formulae for likelihood-based tests of new physics}",
	url = {https://doi.org/10.1140/epjc/s10052-011-1554-0},
	volume = {71},
	year = {2011}}

@article{rankPinto,
    author = "Pinto, Andr\'es and Wu, Zhibo and Balli, Fabrice and Berger, Nicolas and Boonekamp, Maarten and Chapon, \'Emilien and Kawamoto, Tatsuo and Malaescu, Bogdan",
    title = "{Uncertainty components in profile likelihood fits}",
    eprint = "2307.04007",
    archivePrefix = "arXiv",
    primaryClass = "physics.data-an",
    doi = "10.1140/epjc/s10052-024-12877-5",
    journal = "Eur. Phys. J. C",
    volume = "84",
    number = "6",
    pages = "593",
    year = "2024"
}

@article{hifa,
    author = "Cranmer, Kyle and Lewis, George and Moneta, Lorenzo and Shibata, Akira and Verkerke, Wouter",
    collaboration = "ROOT",
    title = "{HistFactory: A tool for creating statistical models for use with RooFit and RooStats}",
    reportNumber = "CERN-OPEN-2012-016",
    journal = "10.17181/CERN-OPEN-2012-016",
    month = "6",
    year = "2012",
    url           = "https://cds.cern.ch/record/1456844",
    doi           = "10.17181/CERN-OPEN-2012-016",
    pages = "{\unskip}",
}

@article{lumiB2,
    author = "Adachi, I. and others",
    collaboration = "Belle~II",
    title = "{Measurement of the integrated luminosity of data samples collected during 2019-2022 by the Belle II Experiment}",
    eprint = "2407.00965",
    archivePrefix = "arXiv",
    primaryClass = "hep-ex",
    reportNumber = "Belle II Preprint 2024-019; KEK Preprint 2024-16",
    doi = "10.1088/1674-1137/ad806c",
    journal = "Chin. Phys. C",
    volume = "49",
    number = "1",
    pages = "013001",
    year = "2025"
}

@article{hepdata,
    author=" ",
    collaboration = "{Belle~II, Belle}",
    title = "{Search for ${B}^0\to {K}_{\rm S}^0\tau^+\tau^-$ decays at Belle and {Belle~II}}",
    journal = "{$\mathrm{HEPData\;(collection)}\!\!$}",
    year = "",
    note="{Link to follow after journal publication.}",
}

@article{LHCb:2025lcw,
    author = "Aaij, Roel and others",
    collaboration = "LHC${\rm b}$",
    title = "{Searches for $B^0\to K^+\pi^-\tau^+\tau^-$ and $B_s^0\to K^+K^-\tau^+\tau^-$ decays}",
    eprint = "2510.13716",
    archivePrefix = "arXiv",
    primaryClass = "hep-ex",
    reportNumber = "LHCb-PAPER-2025-048, CERN-EP-2025-224",
    doi = "10.1103/y968-44v8",
    journal = "Phys. Rev. Lett.",
    volume = "136",
    number = "18",
    pages = "181802",
    year = "2026"
}

@article{bstautau,
    author = "Aaij, Roel and others",
    collaboration = "LHC${\rm b}$",
    title = "{Search for the decays $B_s^0\to\tau^+\tau^-$ and $B^0\to\tau^+\tau^-$}",
    eprint = "1703.02508",
    archivePrefix = "arXiv",
    primaryClass = "hep-ex",
    reportNumber = "CERN-EP-2017-034, LHCB-PAPER-2017-003",
    doi = "10.1103/PhysRevLett.118.251802",
    journal = "Phys. Rev. Lett.",
    volume = "118",
    number = "25",
    pages = "251802",
    year = "2017"
}

@article{Chen:2024jlj,
    author = "Chen, Feng-Zhi and Wen, Qiaoyi and Xu, Fanrong",
    title = "{Correlating $B\rightarrow K^{(*)} \nu \bar{\nu }$ and flavor anomalies in SMEFT}",
    eprint = "2401.11552",
    archivePrefix = "arXiv",
    primaryClass = "hep-ph",
    doi = "10.1140/epjc/s10052-024-13425-x",
    journal = "Eur. Phys. J. C",
    volume = "84",
    pages = "1012",
    year = "2024"
}

@article{Aebischer:2018iyb,
    author = "Aebischer, Jason and Kumar, Jacky and Stangl, Peter and Straub, David M.",
    title = "{A Global Likelihood for Precision Constraints and Flavour Anomalies}",
    eprint = "1810.07698",
    archivePrefix = "arXiv",
    primaryClass = "hep-ph",
    doi = "10.1140/epjc/s10052-019-6977-z",
    journal = "Eur. Phys. J. C",
    volume = "79",
    number = "6",
    pages = "509",
    year = "2019"
}

@article{Gubernari:2023puw,
    author = "Gubernari, Nico and Reboud, M{\'e}ril and van Dyk, Danny and Virto, Javier",
    title = "{Dispersive analysis of $B \to K^{(*)}$ and $B_{s} \to \phi$ form factors}",
    eprint = "2305.06301",
    archivePrefix = "arXiv",
    primaryClass = "hep-ph",
    reportNumber = "EOS-2023-02, IPPP/23/22, P3H-23-026, SI-HEP-2023-09",
    doi = "10.1007/JHEP12(2023)153",
    journal = "JHEP",
    volume = "12",
    pages = "153",
    year = "2023",
    note = "[Erratum: \href{https://doi.org/10.1007/JHEP12(2023)153}{\textit{JHEP} \textbf{01} (2025) 125}]"
}

@article{belle_eid,
title = {Electron identification in {Belle}},
journal = {Nucl. Instrum. Meth. A},
volume = {485},
number = {3},
pages = {490-503},
year = {2002},
issn = {0168-9002},
doi = {https://doi.org/10.1016/S0168-9002(01)02113-1},
url = {https://www.sciencedirect.com/science/article/pii/S0168900201021131},
author = {K Hanagaki and H Kakuno and H Ikeda and T Iijima and T Tsukamoto}
}

@article{belle_muid,
title = {Muon identification in the {Belle} experiment at {KEKB}},
journal = {Nucl. Instrum. Meth. A},
volume = {491},
number = {1},
pages = {69-82},
year = {2002},
issn = {0168-9002},
doi = {https://doi.org/10.1016/S0168-9002(02)01164-6},
url = {https://www.sciencedirect.com/science/article/pii/S0168900202011646},
author = {A Abashian and K Abe and K Abe and P.K Behera and F Handa and T Iijima and Y Inoue and H Miyake and T Nagamine and E Nakano and S Narita and L Piilonen and S Schrenk and Y Teramoto and K Trabelsi and J.G Wang and M Yamaga and A Yamaguchi and Y Yusa}
}

@article{junk_cls,
title = {Confidence level computation for combining searches with small statistics},
journal = {Nucl. Instrum. Meth. A},
volume = {434},
number = {2},
pages = {435-443},
year = {1999},
issn = {0168-9002},
doi = {https://doi.org/10.1016/S0168-9002(99)00498-2},
url = {https://www.sciencedirect.com/science/article/pii/S0168900299004982},
author = {Thomas Junk}
}

@article{Belle:2012yvr,
    author = "Dalseno, J. and others",
    collaboration = "Belle",
    title = "{Measurement of Branching Fraction and First Evidence of CP Violation in $B^0 \to a_1^{\pm}(1260) \pi^\mp$ Decays}",
    eprint = "1205.5957",
    archivePrefix = "arXiv",
    primaryClass = "hep-ex",
    reportNumber = "BELLE-PREPRINT-2012-15, KEK-PREPRINT-2012-7",
    doi = "10.1103/PhysRevD.86.092012",
    journal = "Phys. Rev. D",
    volume = "86",
    pages = "092012",
    year = "2012"
}

@article{lidbelleii,
      title="{Charged-lepton identification at Belle~II}", 
      author={M. Abumusabh and others},
      collaboration = "Belle~II",
      year={2026},
      eprint={2606.26274},
      archivePrefix={arXiv},
      primaryClass={hep-ex},
      url={https://arxiv.org/abs/2606.26274}, 
}

\end{document}